\documentclass[reprint,aps,prd,nofootinbib,a4paper,10pt,superscriptaddress]{revtex4-2}
\usepackage{hyperref}
\usepackage{changes}
\usepackage[section]{placeins}
\usepackage[titletoc]{appendix}
\usepackage{xcolor}
\usepackage{dsfont}
\usepackage{bm}
\usepackage{mathtools}
\usepackage{enumerate}
\usepackage{amsfonts}
\usepackage{lipsum, babel} 
\DeclareMathAlphabet{\mathpzc}{OT1}{pzc}{m}{it} 

\newcommand{\sayy}[1]{`#1'}
\providecommand{\href}[2]{#2}

\makeatletter
\newcommand*{\rom}[1]{\expandafter\@slowromancap\romannumeral #1@}
\makeatother

\def\be{\begin{equation}}
\def\ee{\end{equation}}
\def\bea{\begin{eqnarray}}
\def\eea{\end{eqnarray}}

\def\sig{\sigma}

\def\la{\langle}
\def\ra{\rangle}
\def\Eu{ \mathfrak{H} }

\def\obs{\mathcal{O}}
\def\emi{\mathcal{E}}

\definecolor{MyB}{rgb}{0.1,0.1,1.0}
\newcommand{\eda}[1]{{\color{MyB}{#1}}}

\newcommand{\UD}[2]{\ensuremath{^{#1}_{\phantom{#1} #2}}}
\newcommand{\DU}[2]{\ensuremath{_{#1}^{\phantom{#1} #2}}}

\newcommand{\dd}{\ensuremath{{\rm d}}}

\def\prd{{Phys. Rev.} D }

\def\MNRAS{{Mon.\ Not.\ R.\ Astr.\ Soc.}}

\def\prd{{{Phys.\ Rev. D.}}}

\def\aap{{Astron.\ Astrophys}}

\begin{document}

\title{Exploring the rich geometrical information in cosmic drift signals with covariant cosmography} 
\author{Asta~Heinesen}
\email{asta.heinesen@nbi.ku.dk} 
\affiliation{Niels Bohr Institute, Blegdamsvej 17, 2100 Copenhagen, Denmark} 
\affiliation{Univ Lyon, Ens de Lyon, Univ Lyon1, CNRS, Centre de Recherche Astrophysique de Lyon UMR5574, F--69007, Lyon, France} 

\author{Miko\l{}aj Korzy\'nski} 
\email{korzynski@cft.edu.pl}
\affiliation{Center for Theoretical Physics, Polish Academy of Sciences, Al. Lotnik\'o{}w 32/46, 02-668 Warsaw, 
Poland}

\begin{abstract} 
Real-time measurements are becoming feasible in cosmology, where the next generation of telescopes will detect the temporal change of redshifts and sky positions of individual sources with a precision that will allow a \emph{direct} detection of the cosmic expansion rate. 
These detections of cosmic drifts of redshifts and positions are likely to become  cornerstones in modern cosmology, where one has otherwise relied on the indirect inference of cosmic expansion by estimation of the slope of the fitted distance--redshift relation.  
Because of their ability to directly detect the cosmic time-evolution, real-time measurements are powerful as model-independent probes. 
We develop a cosmographic framework for analysing cosmological redshift drift and position drift signals without knowledge of the space-time geometry. 
The framework can be applied to analyse data from surveys such as the Gaia observatory, the Square Kilometer Array (SKA), and the Extremely Large Telescope (ELT). 
The drift effects are distorted by the regional kinematics and tidal effects in the cosmic neighbourhood of the observer, giving rise to non-trivial corrections to the well known  Friedmann-Lema\^{\i}tre-Robertson-Walker (FLRW) results. We discuss  how one may concretely implement the framework in the statistical analysis of real-time data, along with assumptions and limitations that come with such an analysis. We also discuss the geometrical information that can ideally be extracted from ideal high-resolution data of  cosmic drifts in combination with distance--redshift data.   
\end{abstract}

\maketitle

\section{Introduction}
Real-time observations hold the potential of directly measuring space-time kinematics by following astrophysical sources at cosmic distances over time in the telescope, and registering the corresponding temporal changes in cosmological observables.  
The changes in cosmological observables are also denoted  cosmological drift effects.
The cosmological drifts include the redshift drift \cite{1962ApJ...136..319S,1962ApJ...136..334M,Loeb:1998bu} which will provide an important independent probe of the cosmological expansion rate by upcoming precise measurements by the Extremely Large Telescope (ELT) \cite{Martins:2019gxw} and the Square Kilometer Array (SKA) \cite{Maartens:2015mra}. 
The position drift (or cosmic parallax) can probe large scale structures in the Universe and peculiar acceleration of the observer \cite{Quercellini:2008ty,Krasinski:2010rc,Quercellini:2010zr}. These effects are detectable with the Gaia observatory \cite{2018A&A...616A..14G,2021A&A...649A...9G}. 
Other cosmological drifts of observational interest include the drift of luminosity or angular diameter distance of the source; see \cite{Korzynski:2017nas} and its references for an overview. 
The direct measurements of temporal changes of cosmic redshifts, distances and positions provide a unique opportunity to constrain the kinematics of the Universe model-independently. Loosely formulated, the cosmological drift effects \emph{directly} probe the slope of the Hubble diagram, whereas the conventional measurements of cosmic distances and redshifts rely on the fitting of a cosmological model in order to indirectly infer the slope.

There have been efforts in describing cosmological drift effects while accounting for structures and movements of the emitting sources and the observer, with some of the early progress summarised in \cite{Quercellini:2010zr}. 
In \cite{Liske:2008ph}, emitter motions relative to an idealised Friedmann-Lema\^{\i}tre-Robertson-Walker (FLRW) background have been described, while all first order effects in FLRW perturbation theory were included in \cite{Rasanen:2013swa,Marcori:2018cwn,Bessa:2023qrr}. Drift effects have also been considered for observers in Bianchi I models \cite{Fleury:2014rea,Marcori:2018cwn,Koksbang:2020zej}, off-center observers in Lema\^{\i}tre-Tolman-Bondi models \cite{2010PhRvD..81d3522Q,Krasinski:2010rc}, 2-region models \cite{Koksbang:2023wez}, Newtonian N-body simulations \cite{Koksbang:2023tun}, and  hydrodynamic relativity-simulations \cite{Koksbang:2024xfr}. 
Frameworks for describing cosmological drift effects in general space-time geometries have been considered in \cite{Rasanen:2013swa,Hellaby:2017soj,Korzynski:2017nas,Korzynski:2019oal,Heinesen:2020pms,Heinesen:2021nrc}, along with a computational tool to calculate cosmological drift effects in a given input-space-time \cite{Grasso:2021iwq, PhysRevD.104.043508}. 
In \cite{Heinesen:2021qnl}, cosmography for analysing redshift drift at linear order in the distance to the source in general space-times was formulated. 
The work in \cite{Heinesen:2021qnl} extends the framework for cosmography for distance--redshift data developed in previous analyses, e.g., \cite{Umeh:2013UCT,Clarkson:2011uk,Heinesen:2020bej,Maartens:2023tib}  to include redshift drift, and we will further extend this framework in the present paper.

In this paper, we will explore the potential for performing model-independent constraints based on the cosmological drift effects. We will compute cosmographic series expansions\footnote{Note that the series expansion is in the distance to the source; the space-time metric itself is not assumed to be perturbative around any specified background geometry} of redshift drift and position drift in the geometrical optics approximation -- \emph{without} assuming any model for the space-time geometry of the Universe. The main aim of the paper is to arrive at a cosmographic framework that bears much resemblance to the usual FLRW cosmography, but now with more degrees of freedom to constrain with data due to that the metric is left unspecified.  The idea is, that this framework can be applied directly to perform model-independent analyses of the cosmological drift effects {by a simple substitution of the FLRW cosmography  with the cosmography for a general space-time description.

We start out with outlining the assumptions and limitations of our setup in section \ref{sec:assumptions}. After reviewing  our congruence description for the space-time in section \ref{sec:spacetime}, we formulate the position drift signal in terms of its cosmographic series expansion in section \ref{sec:pdrift}.
In section \ref{sec:zdrift}, we compute the second order cosmography for redshift drift, while building on the first order results derived in \cite{Heinesen:2021qnl}. This allows to consider sources at greater distances than previously possible with the same methods. In section \ref{sec:Observations}, we discuss the practical implementations of the formalism to analyse catalogues of cosmic drift data. 
We provide some errata to existing results in the literature in section~\ref{sec:errata}. We discuss our main results on position drift and redshift drift in section \ref{sec:discussion}. 
With off-set in these main results, we explore the independent information about the large-scale kinematics of matter and the space-time curvature that can be extracted from cosmographic analyses of redshift drift and position drift measurements in section \ref{sec:information}, where we also examine the additional information about the spacetime that can be extracted in combination with measurements of luminosity distance and redshift. 
We conclude in section \ref{sec:conclusion}.

\vspace{4pt}

\noindent
\underbar{Notation and conventions:}
Units are used in which $c=1$. Greek letters $\mu, \nu, \ldots$ label space-time
indices in a general basis, running from 0 to 3. 
The signature of the space-time metric $g_{\mu \nu}$ is $(- + + +)$ and the connection $\nabla_\mu$ is the Levi-Civita connection. 
$R\UD{\mu}{ \nu \sigma \rho}$ denotes the Riemann curvature of the space-time, and $C\UD{\mu}{\nu\sigma\rho}$ is the Weyl curvature tensor. 
Round brackets $(\, )$ containing indices denote symmetrisation in the involved indices and square brackets $[\, ]$ denote anti-symmetrisation. 

In this paper, we will often make use of purely spatial tensors, i.e. tensors orthogonal to a normalized timelike vector $u^\mu$ in all indices, i.e. $T_{\mu\nu}\,u^\mu = T_{\nu\mu}\,u^\mu = 0$. We will still use standard 4-dimensional Greek indices for these objects that are confined to the 3-dimensional space orthogonal to $u^\mu$. We introduce the projector onto this space
\bea
h_{\mu\nu} = g_{\mu\nu} + u_{\mu}\,u_{\nu} \, ,  
\eea
which also serves as the 
spatial 3-metric on the same space. The traceless symmetric part of the purely spatial tensors will be denoted by the brackets
$\langle\rangle$:
\be
T_{\la \mu\nu \ra} = T_{(\mu\nu)} - \frac{1}{3}\,T\UD{\sigma}{\sigma}\,h_{\mu\nu},
\ee
with the coefficient $\frac{1}{3}$ reflecting the effective dimension 3 of the space. This notation will also be used for tensors of higher valence, see \cite{Spencer:1970} and Appendix A of \cite{Heinesen:2020bej} for the traceless decomposition for symmetric and spatially projected tensors.

\section{Assumptions and limitations}
\label{sec:assumptions}

Before we introduce the geometric setup we will spell out the assumptions of the formalism. 
Our assumptions are minimal, as we do not impose any direct}constraints on the space-time metric. 
However, there are certain basic regularity assumptions and conventional simplifying assumptions that we do make.  
\begin{itemize} 
\item  We assume a metric theory of space-time and gravity with a Levi-Civita connection. 
Our results are thus applicable to general relativity as well as modified metric theories of gravity without torsion. 

\item We assume that both the observer and the observed sources (i.e. the stars/galaxies) can be described using a time-like congruence. This means that to each point of the spacetime we assign a single 4-velocity vector describing the cosmic flow of emitters and observers. This assumption obviously involves a coarse-graining step of the complicated small-scale motion of matter. 
The  merging of stars or galaxies or emergence of small-scale virialized structures in the matter distribution thus cannot be accounted for in this formalism. 
Our formalism applies to scales where multivaluedness of 4-velocity is absent or rare, or situations where the effect of this is small enough to be ignored.

\item We assume that the geometrical optics approximation is valid, i.e. light rays propagate along null geodesics of the space-time metric.

\item We further assume that photon number is conserved along the null beams (no loss of photons through particle scattering processes). 
The Etherington distance duality relation between the angular diameter distance $d_A$ and luminosity distance $d_L$ then follows: 
\begin{eqnarray}
    d_{L} = (1+z)^2\,d_A,
\end{eqnarray}
with $z$ being the redshift. 

\item 
In order to make the definition of the drifts well defined for sources in all possible directions on the observer's sky, we must require the null geodesics emanating form the observer not to intersect. In other worlds, we are neglecting caustic behaviour of light. 
This eliminates any kind of strong gravitational lensing effects from the description.

\item Finally, the applicability of the cosmographic framework   
relies on convergence of the Taylor series of the drifts. The framework must be applied in a regime where the series converge and where the truncated Taylor series provide good approximations of the relevant observables, such that any remainder term is reasonably bounded.     
\end{itemize}

\section{The space-time description} 
\label{sec:spacetime}
\begin{figure}
    \includegraphics[width=\columnwidth]{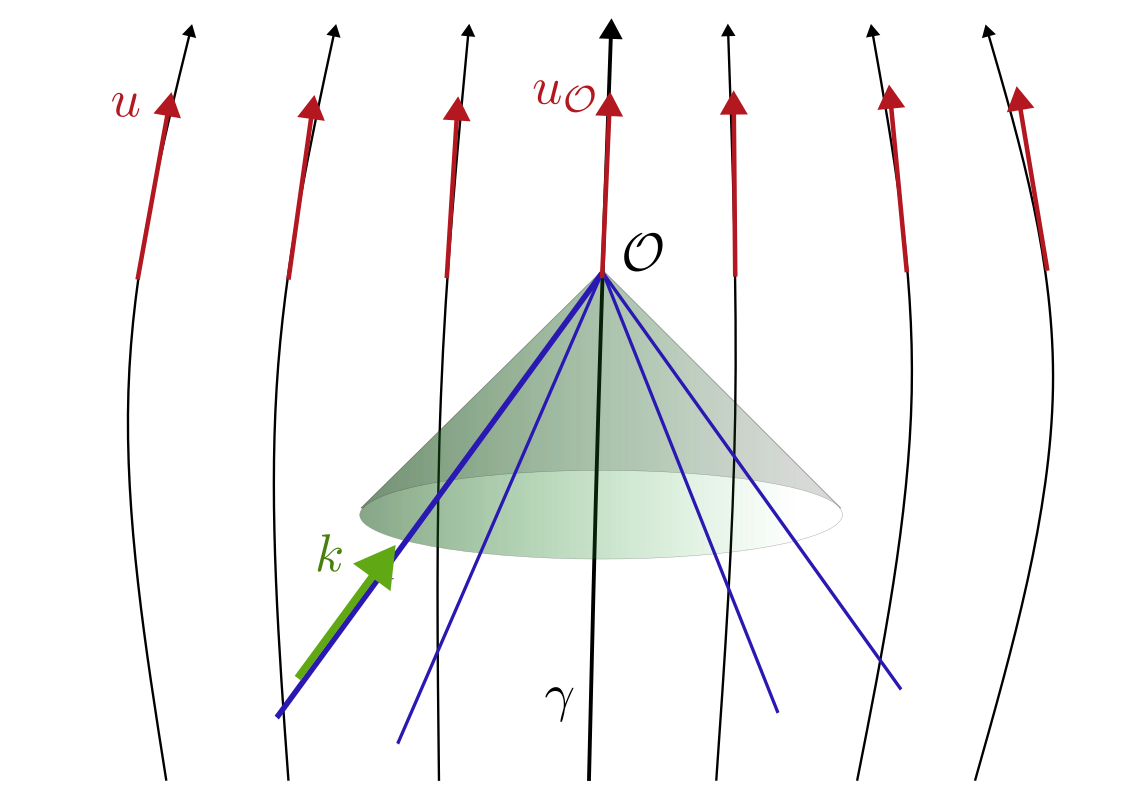}
    \caption{The geometrical setup of the paper: congruence of observers (black), with tangent vector $u^\mu$, and the family of null geodesics (blue) crossing at the observer's worldline $\gamma$, with the tangent vector $k^\mu$. The moment of observation is denoted by $\cal O$.}
    \label{fig:setup}
\end{figure}

We consider a general space-time with the observer and sources belonging to a single time-like congruence (henceforth the \sayy{observer congruence}) with 4-velocity field $u^\mu$, see Figure~\ref{fig:setup}. 
Without loss of generality, we can make the following kinematic decomposition 
\bea
\label{def:expu}
&& \nabla_{\nu}u_\mu  = \frac{1}{3}\theta h_{\mu \nu }+\sig_{\mu \nu} + \omega_{\mu \nu} - u_\nu a_\mu  \ , \nonumber \\  
&& \theta \equiv \nabla_{\mu}u^{\mu} \, ,  \quad \sig_{\mu \nu} \equiv h\DU{ \la \nu  }{ \beta}\,  h\DU{  \mu \ra }{\alpha } \nabla_{ \beta }u_{\alpha  }  \, , \nonumber \\ 
&& \omega_{\mu \nu} \equiv h\DU{  \nu  }{ \beta}\,  h\DU{ \mu }{ \alpha }\nabla_{  [ \beta}u_{\alpha ] }   \, , \quad  a^\mu \equiv \dot{u}^\mu \,  , 
\eea 
where the operator $\dot{} \equiv u^\mu \nabla_\mu$ is the covariant derivative along the worldlines of the observer congruence.

We shall leave the kinematic variables $\theta$, $\sigma_{\mu \nu}$ and $\omega_{\mu \nu}$ completely free in the below analysis. We also keep a general 4-acceleration in this section, and in our results in the appendices, but shall neglect $a^\mu$ in the main results of our paper.  
This is reasonable since non-gravitational interactions are negligible on cosmological scales. 

Let $\gamma$ denote the worldline of the observer and the point ${\cal O}$ on $\gamma$ the point of observation. In order to study drift effects 
of the astrophysical sources, we must consider a segment\footnote{In practice, the proper time interval of this segment can be thought of as the observation time in the observer's telescope.} of $\gamma$ around the point ${\cal O}$ and the light rays emanating from this section. We thus consider a 4-dimensional geodesic congruence of null rays (henceforth the \sayy{photon congruence}) passing from the worldline section of the observer and forming a 1-parameter family of null-cones, one for each instant on the observer's worldline.
We let the future-pointing null vector field $k^\mu$ be the photon 4-momentum of the photon congruence forming the  family of lightcones with vertices along the observer worldline, and let $\lambda$ be the affine parameter along the null geodesic related to $k^\mu$, i.e. $\frac{\dd}{\dd \lambda} = k^\mu\,\partial_\mu$. Moreover, let 
\bea
\label{Ee}
E \equiv - u^\mu k_\mu  \, , 
\eea  
be the photon energy as measured by a member of the observer congruence
  with evolution along the null rays: 
\bea
\label{def:Eevolution}
\frac{ {\rm d} E}{{\rm d} \lambda} = - E^2  \Eu \, , \quad   \Eu \equiv  \frac{1}{3}\theta  - e^\mu a_\mu + e^\mu e^\nu \sigma_{\mu \nu}   \,. 
\eea 
The spatial unit vector $e^\mu \equiv u^\mu - \frac{1}{E} k^\mu$ is the direction of propagation of the photon as seen in the observer congruence frame, and it is by definition orthogonal to $u^\mu$. 
At the an event of observation, $\obs$, the direction vector $e^\mu_\obs$ is multivalued and denotes any direction on the observer's sky. 
However, note that $e^\mu_\obs$ is fixed once we fix a null geodesic (or alternatively fix a source of observation). 
In fact, we may  parametrize all null geodesics emanating from the observer's worldline by the direction $e^\mu_\obs$ and the moment of observation defined by the observers' proper time $\tau$.

From the photon energy in \eqref{Ee}, we define the redshift as measured by the observer at ${\cal O}$ as 
\bea
\label{redshift}
z \equiv \frac{E}{E_\obs} - 1  \, . 
\eea

\section{The position drift signal} 
\label{sec:pdrift} 
\begin{figure}
    \includegraphics[width=\columnwidth]{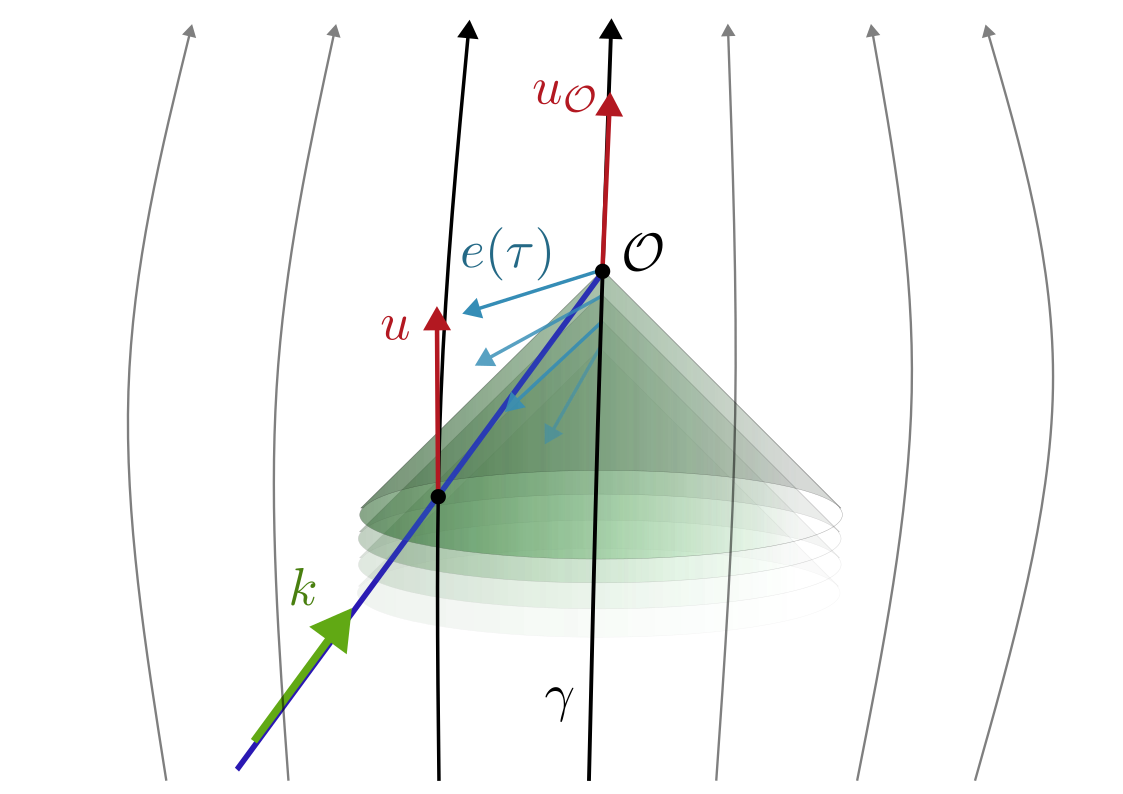}
    \caption{The position drift for a selected source (broader worldline) is the change  of its apparent position vector $e^\mu(\tau)$ as recorded over a period of the observer's proper time $\tau$, defined via the covariant derivative along $\gamma$. Measurement of the drift requires repeated observations of a single source, related to more than one light cone. The position drift, $\kappa^\mu$, at $\cal O$ depends on the momentary velocity of the source, $u^\mu$, as well as the 4-velocity of the observer,  $u^\mu_{\cal O}$. }
    \label{fig:drifts}
\end{figure}
We might consider the drift of photon propagation direction of a given source
\bea
\label{positiondrift}
\kappa^\mu \equiv  h\UD{\mu}{\nu} \,  \dot{e}^\nu \,  \big|_\gamma , 
\eea  
where the source is kept fixed under the time derivative.
In other words, when evaluated at the observer worldline $\gamma$, $e^\mu (\tau)$ in the equation above traces a single light source from the observer congruence for all $\tau$ (moments in observer's proper time), and its covariant derivative gives the momentary drift rate of the source as it appears on the observer's sky, see Figure~\ref{fig:drifts}. The drift is measured with respect to a non-rotating frame defined by the Fermi-Walker transport along the observer's geodesic \cite{Korzynski:2017nas, Grasso:2018mei}. 
Note that for an observer following a geodesic ($a^\mu = 0$) the Fermi-Walker transport actually coincides with the parallel transport.

For real measurements that are corrected for the Earth's motion within the solar system, this geodesic represents the worldline of the Solar System. 
The Solar System, on the other hand, exhibits local motion inside of our Galaxy which again is in motion relative to the center-of-mass of the Local Group of galaxies. These motions are typically thought of as being non-cosmological, and thus, it is often desirable to separate the effects of the motions within gravitationally-bound structures, happening on relatively small scales, from the cosmological effects; see \cite{inoue2019, Marcori:2018cwn} for such efforts. 
We will return to this point of seperation of scales in  Section~\ref{sec:Observations}.

When evaluated at an event at the observer's wordline, $\kappa^\mu_\obs$ is also multivalued and denotes the position drift of the set of astrophysical sources over the observer's sky. 
However, it is single-valued for a given source \cite{Korzynski:2017nas}, under the assumption of no caustics and hence also no multiple imaging. 
Since we have assumed all sources to belong to the observer congruence, \sout{so} their 4-velocity $u^\mu$ is uniquely determined by the source's position. Therefore $\kappa_\obs^\mu$ is unique at a  given moment as a function of the position on the sky given by $e_{\obs}^\mu$, defining a single geodesic, and the distance along the geodesic $\lambda$:
$\kappa_\obs^\mu \equiv \kappa_\obs^\mu(e_\obs^
\nu, \lambda)$. 

Let us now consider the case where the observer and the astrophysical source are close to each other as compared to characteristic curvature scales of the space-time metric. The position drift of the source on the observers sky can then be expanded in terms of the affine distance $\Delta \lambda \equiv \lambda_\emi - \lambda_\obs$ to the source located at the event $\emi$, as detailed in Appendix~\ref{sec:pdriftappendix} using Riemann normal coordinates.  
We write the series expansion of the position drift signal between the source and the observer given in \eqref{positiondrift2} for a non-accelerating observer congruence $a^\mu = 0$ as 
\bea
\label{pdseries}
 \hspace*{-0.6cm}  \kappa^\sigma \rvert_{\obs}  =&&   {}^{(0)} \!{\kappa^\sigma}    \rvert_{\obs}   +    {}^{(1)} \!{\kappa^\sigma}    \rvert_{\obs}   E_\obs \Delta\lambda     +   \mathcal{O}(\Delta \lambda^2)  \, \nonumber \\ 
 =&&   {}^{(0)} \!{\kappa^\sigma}    \rvert_{\obs}   -   \frac{ {}^{(1)} \!{\kappa^\sigma} }{\Eu}   \Bigr\rvert_{\obs}  z     +   \mathcal{O}(z^2) \, , 
\eea  
with 
\bea
\label{pdseriesab}
 \hspace*{-0.05cm}  {}^{(0)} \!{\kappa^\sigma} =&&  p\UD{ \sigma }{ \mu } e^\alpha \,  {}^{(0)}\! \kappa\UD{\mu}{\alpha}   \nonumber \\ 
 \hspace*{-0.05cm}  {}^{(1)} \!{\kappa^\sigma} =&&      \, p\UD{\sigma}{\mu} \Bigl[ \! {}^{(1)}\! \kappa^{\mu}_{0} +  e^\alpha   {}^{(1)}\! \kappa\UD{\mu}{\alpha}   + e^\alpha \! e^\beta \,  {}^{(1)}\! \kappa\UD{\mu}{ \alpha \beta}   
    + \,   e^\alpha \! e^\beta \! e^\gamma  \,  {}^{(1)}\! \kappa\UD{\mu}{\alpha \beta \gamma}   \!  \Bigr] . \nonumber \\ && 
\eea  
Here\eda{,} $p\UD{\mu}{\nu}$ denotes the direction-dependent projection tensor to the screen space, i.e.
\begin{eqnarray}
    p_{\mu\nu} \equiv 
    p_{\mu\nu}(e^\alpha) &=& h_{\mu\nu} - e_\mu\,e_{\nu} \nonumber\\ &=& g_{\mu\nu} + u_\mu\,u_{\nu} - e_\mu\,e_{\nu}
\end{eqnarray}
The coefficient might be computed using \eqref{positiondrift2} and the kinematic decomposition \eqref{def:expu}: 
\bea
\label{pdcoefficients}
 \hspace*{-0.5cm}  {}^{(0)}\! \kappa\UD{\mu}{\alpha}   &&= \sigma\UD{\mu}{\alpha} + \omega\UD{\mu}{\alpha}      \nonumber \\ 
 \hspace*{-0.5cm}  {}^{(1)}\! \kappa^{\mu}_{0}  &&= \frac{1}{6} h^{\alpha \beta} D_{ \alpha}  (\sigma\UD{\mu}{\beta } + \omega\UD{\mu}{\beta })  - \frac{1}{4}    R\UD{\mu}{\nu}  u^\nu   \nonumber \\ 
 \hspace*{-0.5cm}  {}^{(1)}\! \kappa\UD{\mu}{\alpha}  &&= - \theta ( \sigma\UD{\mu}{\alpha} + \omega\UD{\mu}{\alpha}  ) -  (\sigma\UD{\beta}{\alpha} + \omega\UD{\beta}{\alpha})  (\sigma\UD{\mu}{\beta} + \omega\UD{\mu}{\beta})  \nonumber \\   
 \hspace*{-0.5cm} && + \frac{1}{2} R\UD{\mu}{\alpha}  + E\UD{\mu}{\alpha }  - \frac{4}{15} h^{\gamma \beta} (\sigma\UD{\mu}{ \gamma} + \omega\UD{\mu}{ \gamma} ) \sigma_{\alpha \beta }        \nonumber \\    
 \hspace*{-0.5cm}    {}^{(1)}\! \kappa\UD\mu{\alpha \beta}  &&=  \frac{1}{2} D_{\la \alpha}  (\sigma\UD{\mu}{\beta \ra} + \omega\UD{\mu}{\beta \ra})     - \frac{1}{2} C\UD{\mu}{ \alpha \beta \nu} u^\nu   \nonumber \\    
  {}^{(1)}\! \kappa\UD{\mu}{\alpha \beta \gamma} &&=  - (\sigma\UD{\mu}{\la \gamma} + \omega\UD{\mu}{\la \gamma} )  \sigma_{\alpha \beta \ra}  \, ,
\eea 
where $E\UD{\mu}{\alpha} \equiv C\UD{\mu}{\nu \alpha \beta} u^\nu u^\beta$ is the electric Weyl curvature tensor of the space-time. 
The second line of \eqref{pdseries} applies when $\Eu \neq 0$ on the null ray from the emitter to the observer, and follows from the transformation of affine distance to redshift given in \eqref{sec:Jac}.

In the FLRW limit with the observers comoving with the homogeneous and isotropic foliation, we have $\sigma_{\mu \nu} = 0$ and $\omega_{\mu \nu} = 0$. Furthermore, we have that the Weyl tensor vanishes identically, and that the Ricci tensor $R_{\mu \nu}$ has eigen-vector $u^\mu$ along with three degenerate spatial eigenvectors in the plane orthogonal to $u^\mu$. 
This makes the coefficients \eqref{pdseriesab} vanish, and in fact, the position drift \eqref{pdseries} is identically zero at all orders. 
This is expected, due to the isotropy around the observer in the FLRW model. 
Any deviation from isotropy around the observer will generally produce non-zero drifts, and the effected has been calculated for off-center observers in certain Lema\^{\i}tre-Tolman-Bondi models \cite{2010PhRvD..81d3522Q,Krasinski:2010rc}.

\section{The redshift drift signal} 
\label{sec:zdrift} 
The drift of the redshift of an astrophysical source, measured with respect to the observer's proper  time at the event $\obs$,  will be denoted as $\xi_{\cal O} \equiv \left.\frac{dz}{d\tau}\right|_{\cal O}$. It  can be represented by the integral, cf.  \cite{Heinesen:2021qnl}
\bea
\label{redshiftdriftint}
\xi \rvert_{\obs}  \equiv \frac{{\rm d}  z}{{\rm d}  \tau} \Bigr\rvert_{\obs}  = E_\emi \! \! \int_{\lambda_\emi}^{\lambda_\obs} \! \! d \lambda \,   \Pi     \, , \qquad z \equiv \frac{E_\emi}{E_\obs} - 1 \, ,
\eea  
where the affine parameter $\lambda$ of the non-caustic (i.e. non-self-intersecting, except at the observer's worldline) photon congruence satisfies $k^\mu \nabla_\mu \lambda = 1$, and where subscripts $\emi$ and $\obs$ indicate evaluation at the points of emission and observation. 
In the absence of 4-acceleration, such that $a^\mu =0$, the integrand of \eqref{redshiftdriftint} can be expressed as the series (see   \cite{Heinesen:2021qnl} for details of the derivation)
\bea
\label{Pimultikappa}
\hspace*{-0.65cm} \Pi &=&  -  \kappa^\mu \kappa_\mu  + \Sigma^{\it{o}}        +       e^\mu   e^\nu \Sigma^{\bm{ee}}_{\mu \nu} + e^\mu   \kappa^\nu \Sigma^{\bm{e\kappa}}_{\mu \nu}    
\eea  
with coefficients 
\bea
\label{Picoefkappa}
&& \Sigma^{\it{o}} \equiv  - \frac{1}{3} u^\mu u^\nu R_{\mu \nu}     \, , \nonumber  \\   
&& \Sigma^{\bm{ee}}_{\mu \nu} \equiv        -  E_{\mu \nu}   -  \frac{1}{2} h\UD{\alpha}{ \la \mu} h\UD{\beta}{ \nu \ra}  R_{ \alpha \beta }  \,   , \nonumber  \\   
&& \Sigma^{\bm{e\kappa}}_{\mu \nu} \equiv  2 (\sigma_{\mu \nu}  - \omega_{\mu \nu}   )  \,  
\eea  
that are defined over the entire observer congruence, and evaluated at the null ray passing from $\emi$ to $\obs$ for the purpose of calculating the integrand \eqref{Pimultikappa}.  
For sources located close to the observer, we might formally expand the redshift drift signal \eqref{redshiftdriftint} in the separation between the emitter and the observer, measured by the affine parameter difference $\Delta\lambda$  along the null geodesic. 

For the purpose of computing the redshift drift cosmography, it is useful to expand $\Pi$ into its first and second order term\footnote{$\Pi$ as evaluated at the observer is itself dependent on the affine distance to the source through the position drift $\kappa$} in $\Delta \lambda$: 
\bea
\label{Pitaylor}
\hspace{-0.5cm} \Pi  \rvert_{\obs}  =&& {}^{(0)} \!{\Pi} \rvert_{\obs} + {}^{(1)} \!{\Pi} \rvert_{\obs}  E_\obs \Delta\lambda + \mathcal{O}( \Delta\lambda^2 ) \, ,    
\eea  
and similarly for the derivative of $\Pi$: 
\bea
\label{Pidertaylor}
\hspace{-0.5cm} \frac{{\rm d} \Pi }{{\rm d}  \lambda} \Bigr\rvert_{\obs}  =&& {}^{(0)}\!\left(\frac{{\rm d} \Pi }{{\rm d}  \lambda}\right)_{\cal O} + \mathcal{O}( \Delta\lambda ) \, .     
\eea

We may now provide the cosmography for redshift drift in terms of the above-defined expansion coefficients:

\onecolumngrid
\bea
\label{rdtaylor}
\hspace{-0.5cm} \xi\rvert_{\obs} 
&&=  {}^{(1)}\!{\xi} \rvert_{\obs}  E_\obs \Delta\lambda + {}^{(2)}\!{\xi} \rvert_{\obs}  E^2_\obs \Delta\lambda^2 + \mathcal{O}( \Delta\lambda^3 ) \, ;  \qquad  {}^{(1)}\!{\xi}  \equiv -  {}^{(0)}\!{\Pi} \; , \quad  {}^{(2)}\!{\xi}  \equiv -  \left( -\Eu \,\,{}^{(0)}\!{\Pi}  +  \frac{1}{2E}\,\, {}^{(0)}\!\left(\frac{{\rm d} \Pi }{{\rm d}  \lambda}\right) + {}^{(1)}\!{\Pi}  \right)   \nonumber   \\ 
&&=  {}^{(1)}\!{\hat{\xi}} \rvert_{\obs}  z + {}^{(2)}\!{\hat{\xi}} \rvert_{\obs}   z^2 + \mathcal{O}( z^3 ) \, ;  \qquad \qquad \qquad \quad \;\;  {}^{(1)}\!{\hat{\xi}}  \equiv  \frac{ {}^{(0)}\!{\Pi} }{ \Eu } \; , \quad \;\;  {}^{(2)}\!{\hat{\xi}}  \equiv -  \left( \frac{1}{2} \frac{ {}^{(0)}\!{\Pi} }{ \Eu }  + \frac{1}{2E} \frac{ 
{}^{(0)}\!\left(\frac{{\rm d} \Pi }{{\rm d}  \lambda}\right)}{ \Eu^2 } + \frac{ {}^{(1)}\!{\Pi} }{ \Eu^2 } \right) \, ,   
\eea  
where we recall the definition of the operator $\frac{{\dd}  }{{\dd}  \lambda} \equiv k^\mu \nabla_\mu$ as the derivative along the null bundle. 
In order to arrive at a cosmography that can be constrained with data, we decompose the terms of \eqref{rdtaylor} into multipole representations in the direction of the source 
\bea
\label{Pitaylorcoef}
\hspace{-0cm} {}^{(0)}\!{\Pi}  = &&   \Sigma^{\it{o}}  +    e^\mu   e^\nu ( \Sigma^{\bm{ee}}_{\mu \nu} +  {}^{(0)}\! \kappa\UD{\sigma}{\nu}   \Sigma^{\bm{e\kappa}}_{\mu \sigma}   -  {}^{(0)}\! \kappa\UD{\sigma}{\nu}   {}^{(0)}\! \kappa_{\sigma \mu}        )      +  e^\mu e^\nu e^\sigma e^\rho   (- {}^{(0)}\! \kappa_{\mu \nu}   \Sigma^{\bm{e\kappa}}_{\sigma \rho}   + {}^{(0)}\! \kappa_{\mu \nu}  {}^{(0)}\! \kappa_{\sigma \rho}   ) \, ,  \nonumber \\ 
\hspace{-0.5cm}  {}^{(1)}\!{\Pi}   = && - e^\mu ( {}^{(1)}\!   \Sigma^{\bm{e\kappa}}_{\alpha \mu}  + 2  {}^{(0)}\! \kappa_{ \alpha \mu}  )  {}^{(1)}\! \kappa_{0}^{\alpha} - e^\mu e^\nu ( \Sigma^{\bm{e\kappa}}_{\alpha \mu}  + 2  {}^{(0)}\! \kappa_{ \alpha \mu}  )  {}^{(1)}\! \kappa\UD{\alpha}{\nu}      \nonumber \\  
&& - e^\mu e^\nu e^\sigma    ( \Sigma^{\bm{e\kappa}}_{\alpha \mu}  + 2  {}^{(0)}\! \kappa_{ \alpha \mu}  )(  {}^{(1)}\! \kappa\UD{\alpha}{\nu\sigma} + \delta\UD{\alpha}{\nu} {}^{(1)}\! \kappa_{0 \sigma}   )   +  e^\mu e^\nu e^\sigma e^\rho ( \Sigma^{\bm{e\kappa}}_{\alpha \mu}  + 2  {}^{(0)}\! \kappa_{ \alpha \mu}  )  ( \delta\UD{\alpha}{\nu} {}^{(1)}\! \kappa_{\; \sigma \rho}  -  {}^{(1)}\! \kappa\UD{\alpha}{\nu \sigma \rho} )   \nonumber  \\ 
&&+ e^\mu e^\nu e^\sigma e^\rho e^\gamma ( \Sigma^{\bm{e\kappa}}_{\nu \mu}  + 2  {}^{(0)}\! \kappa_{ \nu \mu}  ) \,  {}^{(1)}\! \kappa_{\sigma \rho \gamma}   +  e^\mu e^\nu e^\sigma e^\rho e^\gamma e^\kappa ( \Sigma^{\bm{e\kappa}}_{\nu \mu}  + 2  {}^{(0)}\! \kappa_{ \nu \mu}  )   {}^{(1)}\! \kappa_{\sigma \rho \gamma \kappa} \, \nonumber \\  
 \frac{1}{E} \,\, {}^{(0)}\!\left(\frac{{\rm d} \Pi }{{\rm d}  \lambda}\right) = && \dot{\Sigma}^{\it{o}}   +  e^\mu \left( - D_\mu \Sigma^{\it{o}}  -  ( \Sigma^{\bm{e\kappa}}_{\alpha \mu} - 2 {}^{(0)}\! \kappa_{\alpha \mu} )   \frac{1}{2} R\UD{\alpha}{\gamma} u^\gamma  \right)     \nonumber \\ 
 &&+ e^\mu e^\nu \left( \dot{\Sigma}^{\bm{ee}}_{\mu \nu}  +  {}^{(0)}\! \kappa\UD{\alpha}{\nu} \dot{\Sigma}^{\bm{e\kappa}}_{\mu \alpha} - 2  {}^{(0)}\! \kappa\UD{\alpha}{\mu } \Sigma^{\bm{e e}}_{\alpha \nu}   -  {}^{(0)}\! \kappa\UD{\alpha}{\mu}    {}^{(0)}\! \kappa\UD{\beta}{\nu}  \Sigma^{\bm{e\kappa}}_{\alpha \beta}   + ( \Sigma^{\bm{e\kappa}}_{\alpha \mu} - 2 {}^{(0)}\! \kappa_{\alpha \mu} )  \left(\frac{1}{2} R\UD{\alpha}{\nu}  + E\UD{\alpha}{\nu}    \right)    \right)     \nonumber \\ 
 &&+ e^\mu e^\nu e^\sigma \left( - D_\sigma \Sigma^{\bm{ee}}_{\mu \nu}  -   {}^{(0)}\! \kappa\UD{\alpha}{\nu}  D_\sigma \Sigma^{\bm{e\kappa}}_{\mu \alpha}   +  ( \Sigma^{\bm{e\kappa}}_{\mu\nu} - 2 {}^{(0)}\! \kappa_{\mu\nu} )   \frac{1}{2} R_{\sigma \gamma } u^\gamma  -   ( \Sigma^{\bm{e\kappa}}_{\alpha \mu } - 2 {}^{(0)}\! \kappa_{\alpha \mu } )\,  C\UD{\alpha}{\nu \sigma \gamma} u^\gamma         \right)     \nonumber \\ 
 &&+ e^\mu e^\nu e^\sigma e^\rho \!\! \left( \!  2  \sigma_{\mu \nu}  \Sigma^{\bm{e e}}_{\sigma \rho}    -  {}^{(0)}\! \kappa_{\mu \nu} \dot{\Sigma}^{\bm{e\kappa}}_{\sigma \rho}   +  (\sigma_{\mu \nu}   +  {}^{(0)}\! \kappa_{\mu \nu}     )  {}^{(0)}\! \kappa\UD{\alpha}{\sigma}  \Sigma^{\bm{e\kappa}}_{\rho \alpha}       -   ( \Sigma^{\bm{e\kappa}}_{\mu\nu} - 2 {}^{(0)}\! \kappa_{\mu\nu} )    (\frac{1}{2} R_{ \sigma \rho }  + E_{ \sigma \rho }       )   \right. \nonumber \\  
&& \left.  - \,    \Sigma^{\bm{e\kappa}}_{\mu \nu}    {}^{(0)}\! \kappa_{\alpha \sigma} {}^{(0)}\! \kappa\UD{\alpha}{\rho}  \! \right)     \nonumber \\ 
 &&+ e^\mu e^\nu e^\sigma e^\rho e^\kappa \left(   {}^{(0)}\! \kappa_{\mu \nu}  D_\sigma \Sigma^{\bm{e\kappa}}_{\rho \kappa}   \right)     \nonumber \\ 
 &&+ e^\mu e^\nu e^\sigma e^\rho e^\kappa e^\gamma \left( -  \sigma_{\mu \nu}    {}^{(0)}\! \kappa_{\sigma \rho}  \Sigma^{\bm{e\kappa}}_{\kappa \gamma} +   \Sigma^{\bm{e\kappa}}_{\mu \nu}   {}^{(0)}\! \kappa_{\sigma \rho} {}^{(0)}\! \kappa_{\kappa \gamma}     \right)    \, . 
\eea  
The expressions for ${}^{(0)}\!{\Pi}$ and ${}^{(1)}\!{\Pi}$ can be found by using \eqref{Pimultikappa} and the series expansion \eqref{pdseries}. 
For the purpose of deriving $\frac{1}{E}  {}^{(0)}\!{ \frac{{\rm d}  \Pi }{{\rm d}  \lambda} }$, we have written $\Pi$ in its multipole decomposition \eqref{Pimultikappa} before computing its derivative 
\bea
\label{Pideriv}
\hspace{-0.5cm}  \frac{{\rm d} \Pi }{{\rm d}  \lambda}  =&&        \frac{{\rm d} \Sigma^{\it{o}}   }{{\rm d}  \lambda}         +       e^\mu   e^\nu \frac{{\rm d}  \Sigma^{\bm{ee}}_{\mu \nu} }{{\rm d}  \lambda}  + e^\mu   \kappa^\nu  \frac{{\rm d}   \Sigma^{\bm{e\kappa}}_{\mu \nu}   }{{\rm d}  \lambda}    +    2 \frac{{\rm d} e^\mu  }{{\rm d}  \lambda}    e^\nu \Sigma^{\bm{ee}}_{\mu \nu} +  \frac{{\rm d} e^\mu  }{{\rm d}  \lambda}     \kappa^\nu \Sigma^{\bm{e\kappa}}_{\mu \nu}   - 2  \frac{{\rm d} \kappa^\mu  }{{\rm d}  \lambda} \kappa_\mu +  e^\mu  \frac{{\rm d} \kappa^\nu  }{{\rm d}  \lambda} \Sigma^{\bm{e\kappa}}_{\mu \nu}  \,  . 
\eea  
For each of the terms in \eqref{Pideriv} we compute their multipole expansions with evaluation at $\obs$, 
\bea
\label{Piderivterms}
\frac{1}{E} \frac{{\rm d} \Sigma^{\it{o}}   }{{\rm d}  \lambda}   && =  \dot{\Sigma}^{\it{o}}  - e^\mu D_\mu \Sigma^{\it{o}}     \nonumber \\     
\frac{1}{E} e^\mu   e^\nu \frac{{\rm d}  \Sigma^{\bm{ee}}_{\mu \nu} }{{\rm d}  \lambda}  && =  e^\mu   e^\nu  \dot{\Sigma}^{\bm{ee}}_{\mu \nu}  -  e^\mu   e^\nu e^\sigma D_\sigma \Sigma^{\bm{ee}}_{\mu \nu}   \nonumber \\     \frac{1}{E}   \frac{{\rm d} e^\mu  }{{\rm d}  \lambda}    e^\nu \Sigma^{\bm{ee}}_{\mu \nu} && = - e^\mu   e^\nu   {\,\,}^{(0)}\! \kappa\UD{\alpha}{\mu } \Sigma^{\bm{e e}}_{\alpha \nu}  +   e^\mu   e^\nu  e^\sigma   e^\rho  \sigma_{\mu \nu}  \Sigma^{\bm{e e}}_{\sigma \rho}     \nonumber \\    
\frac{1}{E}  \frac{{\rm d} e^\mu  }{{\rm d}  \lambda}     {\,\,}^{(0)}\!{\kappa^\nu} \Sigma^{\bm{e\kappa}}_{\mu \nu}   && =   - e^\mu   e^\nu   {\,\,}^{(0)}\! \kappa\UD{\alpha}{\mu}    {\,\,}^{(0)}\! \kappa\UD{\beta}{\nu}  \Sigma^{\bm{e\kappa}}_{\alpha \beta}    +      e^\mu   e^\nu e^\sigma e^\rho  (\sigma_{\mu \nu}   +  {}^{(0)}\! \kappa_{\mu \nu}     )  {}^{(0)}\! \kappa\UD{\alpha}{\sigma}  \Sigma^{\bm{e\kappa}}_{\rho \alpha}   -      e^\mu   e^\nu e^\sigma e^\rho e^\kappa e^\gamma  \sigma_{\mu \nu}    {\,\,}^{(0)}\! \kappa_{\sigma \rho}  \Sigma^{\bm{e\kappa}}_{\kappa \gamma}   \nonumber \\   
\frac{1}{E} {\,\,}^{(0)}\!\left({\frac{{\rm d} \kappa^\mu  }{{\rm d}  \lambda}}\right) {\,\,}^{(0)}\!{\kappa_\mu} && = - e^\mu  {\,\,}^{(0)}\! \kappa_{\alpha \mu}   \frac{1}{2} R\UD{\alpha}{ \gamma } u^\gamma +   e^\mu   e^\nu {\,\,}^{(0)}\! \kappa_{\alpha \mu}   \left(\frac{1}{2} R\UD{\alpha}{\nu}  + E\UD{\alpha}{\nu}    \right)     \nonumber \\   
&& \;\;\;\; +  e^\mu   e^\nu e^\sigma \left( {\,\,}^{(0)}\! \kappa_{\mu \nu }  \frac{1}{2} R_{\sigma \gamma } u^\gamma  -  {\,\,}^{(0)}\! \kappa_{\alpha \mu } C\UD{\alpha}{\nu \sigma \gamma} u^\gamma    \right)   - e^\mu   e^\nu e^\sigma e^\rho  {\,\,}^{(0)}\! \kappa_{\mu \nu}   \left(\frac{1}{2} R_{ \sigma \rho }  + E_{ \sigma \rho }    \right)   \nonumber \\   
 \frac{1}{E}  e^\mu  {\,\,}^{(0)}\!\left({\frac{{\rm d} \kappa^\nu  }{{\rm d}  \lambda}}\right) \Sigma^{\bm{e\kappa}}_{\mu \nu}  && =   - e^\mu   \Sigma^{\bm{e\kappa}}_{\alpha \mu}   \frac{1}{2} R\UD{\alpha}{ \gamma } u^\gamma +   e^\mu   e^\nu  \Sigma^{\bm{e\kappa}}_{\alpha \mu}   \left(\frac{1}{2} R\UD{\alpha}{ \nu }  + E\UD{\alpha}{\nu }    \right)     +  e^\mu   e^\nu e^\sigma \left(  \Sigma^{\bm{e\kappa}}_{\mu \nu }  \frac{1}{2} R_{\sigma \gamma } u^\gamma  -   \Sigma^{\bm{e\kappa}}_{\alpha \mu } C\UD{\alpha}{\nu \sigma \gamma} u^\gamma    \right)  \nonumber \\   
&&  \;\;\;\; - e^\mu   e^\nu e^\sigma e^\rho   \Sigma^{\bm{e\kappa}}_{\mu \nu}   \left(\frac{1}{2} R_{ \sigma \rho }  + E_{ \sigma \rho }   +  {}^{(0)}\! \kappa_{\alpha \sigma} {\,\,}^{(0)}\! \kappa\UD{\alpha}{\rho}    \right)  + e^\mu   e^\nu e^\sigma e^\rho e^\kappa e^\gamma   \Sigma^{\bm{e\kappa}}_{\mu \nu}   {}^{(0)}\! \kappa_{\sigma \rho} {\,\,}^{(0)}\! \kappa_{\kappa \gamma}    
\eea  
where all terms involving $\kappa$ and its gradient are evaluated at zeroth order in $\Delta\lambda$. 
In deriving the coefficients we have used \eqref{pdseries} and the result in (\ref{kappaderiv}) for $a^\mu = 0$ which yields  
 \bea
\label{kdotexpand}
\hspace{-0.5cm} && \frac{{\rm d} \kappa^\mu }{ {\rm d} \lambda}  \Big\rvert_{\obs} =  {}^{(0)}\!\left({\frac{{\rm d} \kappa^\mu }{ {\rm d} \lambda}}\right)      \Big\rvert_{\obs}   + \, \mathcal{O}(\Delta\lambda) \,   
\eea 
with 
 \bea
\label{kdotexpand2}
\hspace{-0.5cm} && {}^{(0)}\!\left(\frac{{\rm d} \kappa^\mu }{ {\rm d} \lambda}\right)  =  k^\mu e^\alpha e^\beta  p^{ \sigma  \nu }  {\,\,}^{(0)}\! \kappa_{\sigma \alpha}  {\,\,}^{(0)}\! \kappa_{\nu \beta}       + E\, p\UD{ \mu}{ \nu }  \left[ - \frac{1}{2} R\UD{\nu}{\gamma } u^\gamma  + e^\alpha  \left(\frac{1}{2} R\UD{\nu}{\alpha }  + E\UD{\nu}{\alpha}    \right)  - e^\alpha e^\beta C\UD{\nu}{\alpha \beta \gamma} u^\gamma   \right]    , 
\eea 
together with the identity 
\bea
\label{kderive}
 \frac{{\rm d} e^\mu }{ {\rm d} \lambda} = E\,(e^\mu - u^\mu) \Eu - E\,e^\nu \left(\frac{1}{3}\, \theta h\UD{\mu}{\nu} + \sigma\UD{\mu}{\nu} + \omega\UD{\mu}{\nu} \right) \, . 
\eea  
We can finally add the contributions in \eqref{Piderivterms} to arrive at the result for $\frac{1}{E}  {}^{(0)}\!\left({ \frac{{\rm d}  \Pi }{{\rm d}  \lambda} }\right)$ in \eqref{Pitaylorcoef}. 

The expression for the redshift drift at second order given above may appear complicated. However, we should keep in mind that we are considering a completely general space-time geometry. In light of this, it is remarkable that a cosmographic expression does exist that one can in principle constrain without \emph{a priori} assumptions on the geometry, given a dataset of sufficient quality.   
At leading order in distance from the source, the redshift drift reduces to the rather compact expression  
\bea 
\label{eq:redshiftdrift_of_z}
\xi\rvert_{\obs} = {}^{(1)}\!{\hat{\xi}} \rvert_{\obs}  z  + \mathcal{O}(z^2) =   \frac{ {}^{(0)}\!{\Pi} \rvert_{\obs}  }{\Eu_\obs}  z + \mathcal{O}(z^2)
\eea 
with 
\bea 
\label{eq:Pi0}
{}^{(0)}\!{\Pi}  &=& -\frac{1}{3}\,R_{\alpha\beta}\,u^\alpha\,u^\beta
+\frac{1}{5}\sigma_{\alpha \beta}\,\sigma^{\alpha \beta} + \frac{1}{3}\omega_{\alpha \beta}\,\omega^{\alpha \beta} \nonumber \\ 
&& + \,  e^\alpha e^\beta  \left( - E_{\alpha \beta}  -  \frac{1}{2} h\UD{\mu}{\la \alpha}\, h\UD{\nu}{\beta \ra}  R_{ \mu \nu }  + \frac{3}{7}\sigma\DU{\left\langle \alpha \right.}{\gamma}\,\sigma_{\left.\beta\right\rangle \gamma} - 2\sigma\DU{\left\langle \alpha \right.}{\gamma}\,\omega_{\left.\beta \right\rangle \gamma} + \omega\DU{\left\langle \alpha \right.}{\gamma}\,\omega_{\left.\beta \right\rangle \gamma}\right) - e^\alpha e^\beta e^\gamma e^\sigma \sigma_{\left\langle \alpha \sigma \right.}\,\sigma_{\left.\beta \gamma \right\rangle} 
\eea

\twocolumngrid  

The expression in \eqref{rdtaylor} -- with the multipole decompositons  \eqref{def:Eevolution} and  \eqref{Pitaylorcoef} inserted -- give the final cosmographic expression for the redshift drift to second order. Note that it begins with a linear term, with no constant offset terms present. 
In the FLRW limit, the redshift drift cosmography reduces to the result in equation~(1.3) of \cite{Lobo:2020hcz}. 
Both of the variables $- \frac{ {}^{(0)}\!{\Pi} \rvert_{\obs}  }{\Eu^2_\obs} $ and $\mathfrak{Q}$ reduce to the FLRW deceleration parameter in the FLRW limit, which can be seen by noticing that $\Eu = H$ in this limit, where $H$ is the Hubble parameter, and $\Pi = \Sigma^{\it{o}} = - \ddot{a}a$. The combination $\frac{ {}^{(0)}\!{\Pi} \rvert_{\obs}  }{\Eu^2_\obs}   + {}^{(0)}\!\left({ \frac{{\rm d} \Pi }{{\rm d}  \lambda} }\right)_{\obs}  / ( E_\obs \Eu_\obs^3)$ entering the second order contribution to the redshift drift signal reduces to the FLRW jerk parameter $j \equiv \dddot{a} a^2/ \dot{a}^3$, which can be checked by evaluating this combination while noting that it is only the monopoles $\Sigma^{\it{o}}$ and $\dot{\Sigma}^{\it{o}}$ that can be non-zero.

\section{Observations} 
\label{sec:Observations}
The cosmographic expressions for position drift \eqref{pdseries}--\eqref{pdseriesab} and redshift drift \eqref{rdtaylor}--\eqref{Pitaylorcoef} (alternatively the simplified expression \eqref{eq:redshiftdrift_of_z}-\eqref{eq:Pi0} for leading-order redshift drift cosmography)
can be used to fit datasets of the drift effects with complimentary data of redshift (alternatively angular diameter distance or luminosity distance) and angular position of the sources. 

For redshift drift, this can in practice be done by replacing the FLRW cosmography with the general  cosmography \eqref{rdtaylor}--\eqref{Pitaylorcoef}, and otherwise carrying out the likelihood function  construction as usual. 
The only difference from the FLRW case in terms of implementation is that the data on angular position of the individual sources $e^\mu$ need to be incorporated. 
This is a consequence of abandoning the isotropy ansatz of the FLRW geometry, which naturally leads to higher-order multipoles than the monopole coming into play. 
However, we note that the order of the multipoles to be constrained only go as high as sixth order in $e^\mu$, cf. \eqref{Pitaylorcoef}, thus giving a finite number of degrees of freedom that can in principle be constrained without further reduction by assumptions, given a sufficient quality of data and sky coverage. 

The principle is the same for position drift, where the multipoles entering the cosmography run to third order in $e^\mu$, cf. \eqref{pdseriesab}. 
There is, however, a slight complication in that the position drift signal is a vector quantity, that is effectively 2-dimensional due to the screen space projection with $p\UD{\sigma}{\mu}$, cf. \eqref{pdseriesab}. Thus a vector field with 3 spatial components is projected to a field orthogonal to the radial direction, with 2 independent components at each point of the celestial sphere. 
Since the projection involves loss of information, it is not obvious which combination of terms in the coefficients in \eqref{pdcoefficients} can be recovered from the data. 
A possible way of establishing the degrees of freedom that can be recovered from whole-sky position drift measurements is to make use of the vector spherical harmonic (VSH) decomposition of $\kappa^\mu$, see for example \cite{2012A&A...547A..59M, RGBarrera_1985} for the mathematical details, and \cite{Marcori:2018cwn} for application to the position drift signal\footnote{Vector fields transverse to the radial directions can be decomposed into poloidal and toroidal VSHs. This decomposition is equivalent to expressing the total drift vector as a sum of a gradient term (poloidal) and a curl term (toroidal), both given by scalar functions on the celestial sphere.}.  In this context, it can be shown that the shear $\sigma_{\mu\nu}$ 
can in fact be recovered completely from the poloidal quadrupole, while the vorticity $\omega_{\mu\nu}$ is contained in the toroidal dipole, see Appendix~\ref{ap:vsh}.

As in all cosmological analysis, there is the difficulty in seperating physics coming from small and intermediate scales from the large scale cosmological imprints in the signal. 
This is true in the cosmographic analysis as well, where the scale of the dataset under consideration and the truncation of the Taylor series expansion of the observable (position drift and redshift drift in our case) implicitly imposes a choice of smoothing scale. When the objective is to infer cosmological information from datasets probing scales well above scales of the largest gravitationally-bound structures, it becomes crucial to account for systematic imprints from peculiar motion within bound structures, including the Newtonian 3-acceleration, of the observer \cite{inoue2019, Marcori:2018cwn}. 
This can be done in a seperation-of-scales approach. To leading order, the peculiar imprints on the signal will add to the above-derived results for a large scale congruence in exactly the same way as described for the FLRW scenario \cite{inoue2019, Marcori:2018cwn}. This means that there will be a peculiar poloidal dipole signal in the position drift (the aberration drift signal due to the Galactic and extragalactic accelerations, measured in \cite{Xu2012, titov2011, gaia2021}) and 
an associated peculiar
dipole in the redshift drift, both pointing in the same direction. Due to their dipolar signatures these effects will not mix with the cosmological signal from $\sigma_{\mu\nu}$ and $\omega_{\mu\nu}$, contained in different multipole moments. We leave a full treatment of peculiar small-scale effects to future work. 

We should keep in mind that for the position drift signal, one must account for rotational degrees of freedom relating to the Sun's motion within our Galaxy and our Galaxy's motion within the Local Group in order to make reliable inferences of $\omega_{\mu \nu}$ at larger scales. 
This amounts to determining an appropriate irrotational coordinate frame along the observer's geodesic, valid on larger scales.

 A non-rotating frame for the purpose of astronomy can be operationally defined with the help of distant quasars, whose positions are treated as fixed \cite{2018A&A...616A..14G}. 
This, however, does not neccessary correspond to the desired reference frame for analysing the measured position drift. Recall that the drift in (\ref{positiondrift}) is defined with respect to the  Fermi-Walker-transported   frame, defined at the scale of the observer congruence. 
While this type of frame is easy to consider theoretically for a given fixed observer congruence, it is fairly difficult to give it an operational definition in the real Universe. This is in part because the scale that one considers \sayy{cosmological}, i.e., the border between \sayy{local} astrophysics and \sayy{global} cosmology is not \emph{a priori} defined and must be specified for the problem at hand. 
For most cosmological data analyses, the scale of interest is above that of the largest bound structures, and one therefore needs to correct for any motion/geometrical effects below such scales. 
One might attempt to model the mechanics of the Solar System within the Galaxy, the motion of the Galaxy within the Local Group, etc. This however adds an additional layer of complication to the problem. 
In practice the coarse-graining scale for the cosmological congruence that is probed in a given cosmogaphic analysis is set by the survey geometry and the order of truncation of the series expansion of the observable. As a general rule, the deeper the survey is in terms of radial covarage, and the lower order of truncation of the Taylor series for the observable, the larger is the effective coarsegraning scale for the cosmological congruence that can be inferred.

A simple way out of the problem of determining the irrotational reference frame would be to \emph{postulate $\omega_{\mu \nu}$ to be negligible on the cosmological distances probed}. This is usually assumed, since the vector mode in $\Lambda$CDM cosmology decrease quickly with scale. 
However, assumptions of this kind always come with a price. Namely, we must sacrifice some of the generality of the cosmographic approach. In particular, this assumption is obviously incompatible with metric models such as the G\"odel universe. Still, the resulting space of irrotational congruence descriptions would be much larger than those  of the FLRW class of models.

\section{Errata to previous papers} 
\label{sec:errata}
Here, we list a number of errors/typos that appeared in previous papers by AH on cosmography. 
Equation (2.7) in \cite{Heinesen:2020bej} has wrong sign of the term $-\frac{1}{2} R_{\la \mu \nu \ra} := -\frac{1}{2} h\UD{\alpha}{\la \mu}\, h\UD{\beta}{\nu \ra} \,R_{\alpha \beta}$, where \sayy{$-$} should be replaced by \sayy{$+$}. 
This error propagates to equation (B.2) of the same paper, where $\overset{2}{\mathfrak{r}}_{\mu \nu}$, where $+\frac{1}{2} R_{\la \mu \nu \ra}$ should be replaced by $-\frac{1}{2} R_{\la \mu \nu \ra}$. 
The same error is present in \cite{Heinesen:2021qnl}, where $-\frac{1}{2} h\UD{\alpha}{\la \mu}\, h\UD{\beta}{\nu \ra}\, R_{\alpha \beta}$ should be replaced by $+\frac{1}{2} h\UD{\alpha}{\la \mu}\,h\UD{\beta}{\nu \ra}\,R_{\alpha \beta}$ in the equation for $\Sigma^{\bm e \bm e}_{\mu \nu}$ in Equation (10) and Equation (17). 

There is an additional sign error in Equation (4.3) of \citep{Heinesen:2020bej}, where the term $+a^\nu \omega_{\mu \nu}$ should be replaced by $-a^\nu \omega_{\mu \nu}$ in the formula for $\overset{1}{\mathfrak{q}}_\mu$. 

The sign errors do not have consequences for the results of the application studies carried out so far in, for instance, \citep{Macpherson:2021gbh,Heinesen:2021azp,Dhawan:2022lze,Macpherson:2022eve,Koksbang:2022upf,Cowell:2022ehf}, where anisotropic stress and vorticity were assumed to be zero/subdominant in the numerical simulations and observational schemes employed in the respective works.

\section{Discussion of the results} 
\label{sec:discussion} 
The main results of our paper are the cosmogaphic expressions for position drift and redshift drift. 
The cosmographic Taylor series expansions {of redshift drift and position drift, consituting the main result of this paper, have been calculated by hand independently by both authors and verified by a Mathematica-based computer code with the computer tensor algebra package xTensor and xPerm, part of the xAct package \cite{xAct, xTensor, MARTINGARCIA2008597}. 

The results for positon drift are given in \eqref{pdseries} -- together with the multipole decompositions of ${}^{(0)}\!{\kappa^\sigma}$ and ${}^{(1)}\!{\kappa^\sigma}$, obtained by combining \eqref{pdseriesab} and \eqref{pdcoefficients} -- which provides the expression for position drift in any geometry for which the assumptions discussed in \ref{sec:assumptions}. 
The result for redshift drift is given by  \eqref{rdtaylor} -- together with the multipole decompositions of $\Eu$, ${}^{(0)}\!{\Pi}$, ${}^{(1)}\!{\Pi}$, and ${}^{(0)}\!{ \frac{{\rm d}  \Pi }{{\rm d}  \lambda} }$ given in \eqref{def:Eevolution} and \eqref{Pitaylorcoef} -- which provides the general geometrical expression of redshift drift to second order in the redshift (or alternatively the affine distance). 
If we consider only the leading order term in redshift, the results greatly simplify and can be written more compactly as in \eqref{eq:redshiftdrift_of_z}-\eqref{eq:Pi0}.}  

These results, although complicated looking with all of their terms, are remarkable: 
they show that it is possible to predict the position and redshift drift signals in a given direction on the observer's sky for a close-by object in terms of a finite number of physically interpretable geometric variables. 
Assuming that that certain combinations of kinematic variables of the observer congruence and curvature variables of the space-time are known, the drift signal and position drift for a sufficiently close-by source in a given direction can be predicted. 
Conversely, it is possible to \emph{extract} information of these combined space-time variables given sufficient amount of measurements of redshift drift and position drift over the observer's sky. 
This in turn allows us to draw important information about the Universe model-independently.

One of the important 
implications of this paper\sout{,} is that it will be possible to constrain more kinematic and curvature degrees-of-freedom than what is possible with cosmographic analysis of distance--redshift data alone. 
We will present details of the possibilities for deriving combined constraints using conventional distance--redshift data and the novel cosmic drift measurements in Section~\ref{sec:information}. Before that, we recall the cosmography results for the generalized Hubble law. 
The second order cosmography for luminosity distance, $d_L$, reads\footnote{See \cite{Heinesen:2020bej} for the expression for the third order term in the cosmography, and see 
 \cite{adamek2024cosmography} for the angular diameter distance and its inverse series with redshift as parametrised in terms of distance. } \cite{Umeh:2013UCT,Heinesen:2020bej}
\begin{equation}\label{eq:series}
    d_L(z,\boldsymbol{e}) =  \frac{1}{\Eu_o (\boldsymbol{e}) } \, z   + \frac{1 - \mathfrak{Q}_o(\boldsymbol{e}) }{2 \Eu_o(\boldsymbol{e})} (\boldsymbol{e})\,z^2  + \mathcal{O}( z^3),
\end{equation} 
where the \emph{effective deceleration parameter} can be written in terms of the multipole decompositon
\begin{equation}
\begin{aligned}
\label{q}
\mathfrak{Q}(\boldsymbol{e} ) &= - 1 -  \frac{1}{\Eu^2(\boldsymbol{e} )} \bigg(
\overset{0}{\mathfrak{q}}   +  e^\mu  \overset{1}{\mathfrak{q}}_\mu   +    e^\mu e^\nu   \overset{2}{\mathfrak{q}}_{\mu \nu}     \\ &+   e^\mu e^\nu e^\rho \overset{3}{\mathfrak{q}}_{\mu \nu \rho}    +   e^\mu e^\nu e^\rho e^\kappa  \overset{4}{\mathfrak{q}}_{\mu \nu \rho \kappa} \bigg)     \, , 
\end{aligned}
\end{equation}
with coefficients 
\bea
\label{qpoles}
&& \overset{0}{\mathfrak{q}} \equiv  \frac{1}{3}   \frac{ {\rm d}  \theta}{{\rm d} \tau} + \frac{1}{3} D_{   \mu} a^{\mu  } - \frac{2}{3}a^{\mu} a_{\mu}    - \frac{2}{5} \sigma_{\mu \nu} \sigma^{\mu \nu}    \, , \nonumber \\ 
&& \overset{1}{\mathfrak{q}}_\mu \equiv  - \frac{1}{3} D_{\mu} \theta   -  \frac{2}{5}   D_{  \nu} \sigma\UD{\nu }{ \mu  }   -  \frac{ {\rm d}  a_\mu }{{\rm d} \tau}  + a^\nu \omega_{\mu \nu}  +  \frac{9}{5}  a^\nu \sigma_{\mu \nu}     \, , \nonumber \\
&& \overset{2}{\mathfrak{q}}_{\mu \nu}  \equiv     \frac{ {\rm d}  \sigma_{\mu \nu}   }{{\rm d} \tau} +   D_{  \la \mu} a_{\nu \ra } + a_{\la \mu}a_{\nu \ra }     - 2 \sigma_{\alpha (  \mu} \omega\UD{\alpha}{ \nu )}   - \frac{6}{7} \sigma_{\alpha \la \mu} \sigma\UD{\alpha}{ \nu \ra }   \, , \nonumber \\ 
 && \overset{3}{\mathfrak{q}}_{\mu \nu \rho}  \equiv -  D_{ \la \mu} \sigma_{\nu   \rho \ra }    -  3  a_{ \la \mu} \sigma_{\nu \rho \ra }    \, , \nonumber\\
 && \overset{4}{\mathfrak{q}}_{\mu \nu \rho \kappa}  \equiv  2   \sigma_{\la \mu \nu } \sigma_{\rho \kappa \ra} \, . 
\eea 
Together with the multipole decompostion of $\Eu$ in \eqref{def:Eevolution}, this provides the full multipole decomposition of the generalized Hubble law at second order.

\section{Methods of direct measurements of kinematical quantitites and curvature}
\label{sec:information}

Assume we managed to perform the measurements of the multipoles of the leading order 
 generalized Hubble  law, i.e. $d_L(z,e^i)$ relation \eqref{eq:series}, the position drift \eqref{pdseries} and the redshift drift \eqref{rdtaylor}. 
The monopole and the quadrupole part of the Hubble law gives $\theta$ and $\sigma_{\mu\nu}$ respectively\eda{;} see \eqref{eq:series} and the definition of $\Eu$ in \eqref{def:Eevolution}. Additionally, the dipole component of $\Eu$ gives the large-scale, 4-acceleration $a_{\mu}$, if we allow for acceleration of this kind.

From (\ref{pdseries})-(\ref{pdcoefficients}) we see that the measurement of the leading order position drift $p\UD{\sigma}{\mu}\,\,{}^{(0)}\!\kappa\UD{\mu}{\alpha}\,e^\alpha$ gives
$\sigma_{\mu\nu}$ as well as $\omega_{\mu\nu}$. The measurement of the latter, as we mentioned in Section~\ref{sec:Observations}, is possible provided that we have a non-rotating reference frame at our disposal. 

On top of that, as we mentioned in Section~\ref{sec:Observations}, the three components of the large-scale non-gravitational acceleration $a^\mu$  are contained in the dipole, although we dropped this term in our results for the coefficients \eqref{pdcoefficients}.  

Since the shear $\sigma_{\mu\nu}$ can be obtained from both the position drift and the generalized Hubble law, it is in principle possible to perform a consistency check of the results, since datasets probing the same scales of an underlying cosmological congruence should yield the same results.
In practice, it requires some care to compare kinematic quantities extracted with cosmographic methods from different surveys. This is because the survey geometry and the order of the cosmographic expressions (i.e. the order of truncation of the series expansion applied) will implicitly impose a coarse-graining scale on the derived kinematic quantities. It is for instance obvious that the effective Hubble law as truncated at linear order and applied to datapoints with redshifts of $\gtrsim 0.1$ will effectively coarse-grain over shear and differential expansion at scales corresponding to redshifts $\ll 0.1$, i.e., the $\theta$ and $\sigma_{\mu \nu}$ inferred would have the interpretation as being variables in a congruence description applying at scales with $z \sim 0.1$. Such results would therefore not be quantitatively comparable with other cosmography results for surveys probing scales of $z \ll 0.1$ or $z \gg 0.1$.

With the above caveats in mind, let us assume that we have performed complementary measurements of position drift, redshift drift, and the distance--redshift relation on scales that are comparable, and that we have applied the leading order cosmographic expressions to these measurements. This means that, in particular, we have measured the direction-dependent Hubble parameter $\frak{H}_{\cal O}$ at the probed scale. 
We can subsequently consider what kind of new information can be obtained from the measurement of redshift drift at leading order, given by
${}^{(1)}\!\hat\xi\big|_{\cal O}$ from
\eqref{eq:redshiftdrift_of_z}-\eqref{eq:Pi0}.

The monopole part of the product ${}^{(1)}\!\hat\xi\big|_{\cal O}\,\mathfrak{H}_{\cal O}$ contains 
a combination of the vorticity $\omega_{\mu\nu}$, the shear $\sigma_{\mu\nu}$ and the Ricci tensor component $R_{\alpha\beta}\,u^\alpha u^\beta$.
Therefore, assuming that we have already measured or constrained $\omega_{\alpha\beta}$ and $\sigma_{\alpha\beta}$, we can obtain the first curvature component $R_{\alpha\beta}\,u^\alpha u^\beta$ from the monopole part of the product ${}^{(1)}\hat\xi\big\lvert_{\cal O}\,{\mathfrak{H}}_{\cal O}$. In a similar way the quadrupole of the product yields the combination curvature components $E_{\alpha\beta} + \frac{1}{2}\,R_{\mu\nu} \,h\UD{\mu}{\langle\alpha}\,h\UD{\nu}{\beta\rangle}$. Finally, the highest spherical harmonic $l=4$ yields the product of two shears. This provides another opportunity for a consistency test of the shear measurement.

\section{Conclusion} 
\label{sec:conclusion} 
We have examined cosmographic frameworks relevant for deriving kinematic and curvature variables directly from cosmological data, without making assumptions about the metric of the Universe beforehand. 
These frameworks have the potential to be applied to real data in much the same way as FLRW cosmography, but with additional variables to be constrained. 
The prospects of doing so with distance--redshift data has already been demonstrated with numerical simulations \cite{adamek2024cosmography} and a few first applications to data from supernovae of type 1a \cite{Dhawan:2022lze,Cowell:2022ehf}. 

In this paper we have focused on measurements of cosmological position drifts and redshift drifts by, for instance, the Gaia observatory and the upcoming ELT and SKA facilities. 
The cosmological constraints derivable from such measurements will complement constraints coming from the distance--redshift data. In particular, applying the cosmographic formalism of this paper will allow to constrain new kinematic and curvature variables. 
In a strict FLRW universe, the only multipoles that remain are the monopole components of the observational signals. In perturbed scenarios, the higher order multipoles can however be important, particularly when the distances to the observed sources are modest. The expected anisotropy in the redshift drift signal over the redshift range probed by SKA may be substantial  \cite{Koksbang:2024xfr}. 
The results in this paper allow for a consistent observational treatment of such anisotropies, while at the same time allowing to remain rather agnostic about the space-time geometry (in $\Lambda$CDM cosmology we may think of this as remaining agnostic towards the types and amplitudes of the perturbations). 
As we have detailed in \ref{sec:Observations}, the general cosmographic expressions derived in this paper can be applied to data much in the same way that the special case of the FLRW cosmography is currently applied, by simply substituting the geometrical prediction of the FLRW cosmography (for instance for redshift drift) with the generalised cosmography provided in this paper.

\vspace{6pt} 
\begin{acknowledgments}
AH is funded by the Carlsberg foundation. 
This work is part of a project that has received funding from the European Research Council (ERC) under the European Union's Horizon 2020 research and innovation programme (grant agreement ERC advanced grant 740021--ARTHUS, PI: Thomas Buchert).  
\end{acknowledgments}

\appendix

\section{General results in Riemann normal coordinates}
\label{sec:rnc}
In Riemann normal coordinates adapted to the point $\obs$, we have the following results 
\bea
\label{metric}
&& g_{\alpha \beta} = \eta_{\alpha \beta} - \frac{1}{3} R_{\alpha \mu \beta \nu} \rvert_{\obs} \, x^\mu x^\nu + \mathcal{O}(x^3)
\eea 
with $x^\mu_\obs \equiv 0$, and where $\eta_{\alpha \beta}$ is the Minkowski metric. 
The Christoffel symbols read 
\bea
\label{christoffel}
&& \Gamma\UD{\alpha}{\beta \gamma} =  - \frac{1}{2} ( R\UD{\alpha}{ \beta \gamma \mu}  + R\UD{\alpha}{  \gamma  \beta \mu} ) \rvert_{\obs} \, x^\mu + \mathcal{O}(x^2) \, . 
\eea 
For any geodesic as generated by $k^\mu$ with affine parameter $\lambda$ and passing through the point $\obs$, we have 
\bea
\label{x}
&& x^\mu = k^\mu   \rvert_{\obs} \,\Delta\lambda   + \mathcal{O}(\Delta\lambda^3) \, , 
\eea 
where $\Delta\lambda \equiv \lambda - \lambda_\obs$, with $\lambda$ defined through the transport rule $k^\mu \nabla_\mu \lambda = 1$. 
For any vector field, $V^\mu$, expanded around the point $\obs$, we have 
\bea
\label{vector} 
 V^\mu &=& V^\mu  \rvert_{\obs}  + \nabla_\alpha V^\mu \rvert_{\obs} \,x^\alpha \nonumber  \\
 &+&\frac{1}{2} ( \nabla_\alpha \! \nabla_\beta V^\mu  + R\UD{\mu}{ \alpha  \beta \nu} V^\nu  ) \rvert_{\obs} \,x^\alpha  x^\beta  \nonumber \\
&=&  V^\mu  \rvert_{\obs}  + k^\alpha \nabla_\alpha V^\mu \rvert_{\obs} \,\Delta\lambda   \nonumber \\ 
&+&   \frac{1}{2} k^\alpha k^\beta ( \nabla_\alpha \! \nabla_\beta V^\mu  + R\UD{\mu}{ \alpha  \beta \nu} V^\nu  ) \rvert_{\obs} \,\Delta\lambda^2  \nonumber \\
&+& \mathcal{O}(\Delta\lambda^3) , 
\eea 
where the last equality follows from \eqref{x}. 

\section{Transformation from affine distance to the redshift} 
\label{sec:Jac}
When the energy function is monotonic along the null rays, i.e., $\Eu \neq 0$ on the null ray from the emitter to the observer, we can change expansion variable from affine distance, $\lambda$, to 
redshift, $z$, by using 
\be
\label{lambdaexpand}
\Delta \lambda = \frac{ \partial  }{ \partial z } \bigr\rvert_{  \bm{k}  }    \lambda  \Bigr\rvert_{\obs} z +  \frac{1}{2}  \frac{ \partial^2  }{ \partial z^2 } \bigr\rvert_{  \bm{k}  }    \lambda  \Bigr\rvert_{\obs} z^2  +  \mathcal{O}( z^3)  \ ,
\ee 
where $\bigr\rvert_{  \bm{k}}$ 
after the differential operator
denotes differentiation along the null geodesic with tangent vector $k^\mu$. 
The coefficients  yield
 \cite{Heinesen:2020bej}
\bea
\label{lambdacoef}
\hspace*{-0.6cm} \frac{ \partial  }{ \partial z } \bigr\rvert_{  \bm{k}  }    \lambda  \Bigr\rvert_{\obs} &=&   - \frac{1}{E_{\obs} \Eu_{\obs}} \, , \nonumber  \\
\hspace*{-1cm} \frac{ \partial^2  }{ \partial z^2 } \bigr\rvert_{  \bm{k}  }    \lambda  \Bigr\rvert_{\obs} &=&    \frac{1}{E_{\obs} \Eu_{\obs}}  \left( 3 + \mathfrak{Q}_{\obs}    \right) , \quad \mathfrak{Q} \equiv - 1 -  \frac{1}{E} \frac{     \frac{ {\rm d} \Eu}{{\rm d} \lambda}    }{\Eu^2}  . 
\eea  
It furthermore follows from Sachs optical equations for the angular diameter distance, $d_A$, that 
\be
\label{lambdaexpanddA}
\Delta \lambda = - \frac{1}{E_\obs} d_A  +  \mathcal{O}( d_A^3)  \ ,
\ee 
which makes substitution of affine distance with angular diameter distance trivial up to second order.

\section{Series expansion of position drift}
\label{sec:pdriftappendix} 
The method of obtaining the position drift is based on the geodesic deviation equation for null geodesics and has been described in \cite{Korzynski:2017nas, Grasso:2018mei}. The gist of the method is to derive the deviation vector $X^\mu$ along a null geodesic connecting the observer and the source, corresponding to the null geodesic connecting them at a slightly later moment. Here we employ this method to derive the first order Taylor series expansion of position drift as a function of the affine distance to the source. 

\begin{figure}
    \includegraphics[width=\columnwidth]{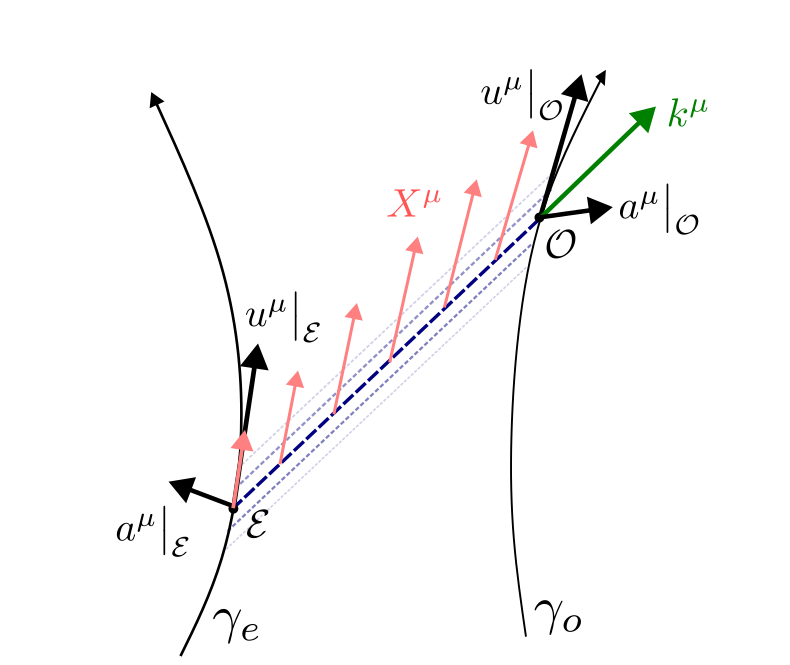}
    \caption{Congruence of null geodesic connecting the worldlines $\gamma_o$ and $\gamma_e$. The central null ray connects $\cal E$ and $\cal O$ and its tangent is $k^\mu$. The deviation vector $X^\mu$, corresponding to a geodesic infinitesimally close to the central one, coincides with $u^\mu$ at $\cal O$ and it is proportional to $u^\mu$ at $\cal E$. $a^\mu$ denotes the acceleration vector of both the emitter and the observer. }
    \label{fig:two-worldlines}
\end{figure}

Let us consider a one-parameter null congruence connecting the observer worldline, $\gamma_o$, and the worldline of an emitter, $\gamma_e$, with the intersection with the emitter's worldline $\gamma_e$ lying in the past of the intersection with the observer's worldline.
Let $k^\mu$ be the 4-momentum of the photon congruence with a central null ray $\gamma_{\bm{k}}$ connecting the points $\obs$ on $\gamma_o$  and $\emi$ on $\gamma_e$ respectively. Let $u^\mu$ be a twice differentiable vector field such that $u^\mu \rvert_{\gamma_o}$ is the observer 4-velocity, $u^\mu \rvert_{\gamma_e}$ is the emitter 4-velocity, and $u^\mu \rvert_{\gamma_{\bm{k}}}$ is any smooth extension of these 4-velocities along the central null ray, see Figure~\ref{fig:two-worldlines}.
 We define the position drift of the emitter as viewed by the observer as  
\bea
\label{positiondriftapp}
\kappa^\mu \rvert_{\obs}  &\equiv&  p\UD{\mu}{\nu }  u^\alpha \nabla_\alpha e^\nu \rvert_{\obs} =   p\UD{ \mu }{ \nu }  a^\nu \rvert_{\obs} - \frac{1}{E} p\UD{ \mu }{ \nu }  u^\alpha \nabla_\alpha k^\nu \rvert_{\obs}  \nonumber \\ 
 &=& p\UD{ \mu }{ \nu }\,  a^\nu \rvert_{\obs} - \frac{1}{E} p\UD{ \mu }{ \nu }\,  k^\alpha \nabla_\alpha X^\nu  \rvert_{\obs}  \, , 
\eea
where $e^\mu \rvert_{\gamma_o}$ is pointing towards the spatial direction of the incoming photon as viewed by the observer, and is defined through the decomposition $k^\mu = E  (u^\mu  - e^\mu)$ as evaluated on the observer worldline. 
Recall that the projector $p\UD{ \mu }{ \nu } = u^{\mu}u_\nu - e^{\mu} e_\nu + \delta\UD{ \mu }{ \nu }$ as defined on the observer worldline gives the projection of tensors onto the angular plane transverse to the incoming photon. 
The deviation vector $X^\mu$ is defined from a vanishing lie derivative along $k^\mu$, such that: $k^\mu \nabla_\mu X^\nu  =  X^\mu \nabla_\mu k^\nu$. Furthermore we require 
\bea
\label{deviationini}
X^\mu \rvert_{\obs} = u^\mu \rvert_{\obs} \, , \qquad X^\mu \rvert_{\emi} = E_\obs u^\mu \rvert_{\emi}/E_\emi
\eea
such that $X^\mu$ is tangent to the emitter and observer worldlines, thus producing the congruence between them of interest. 
Note that the prefactor $E_\obs / E_\emi = \frac{1}{1+z}$ in front of $u^\mu\big|_\emi$ is necessary to ensure consistency with the requirement that $k^\mu$ is a parallel transported null vector; see \cite{Korzynski:2017nas, Grasso:2018mei}. 
For an emitter that is close to the observer, we have 

\onecolumngrid

\bea
\label{deviation}
X^\mu \rvert_{\emi} = X^\mu \rvert_{\obs} + k^\alpha \nabla_\alpha X^\mu \rvert_{\obs} \Delta\lambda + R\UD{\mu}{ \alpha  \beta \nu} k^\alpha k^\beta  X^\nu   \rvert_{\obs} \,\Delta\lambda^2   + \mathcal{O}(\Delta\lambda^3) , 
\eea
where we have used \eqref{vector} and the geodesic deviation equation 
\bea
\label{Gdeviation}
k^\alpha k^\beta \,\nabla_\alpha \! \nabla_\beta X^\mu =  R\UD{\mu}{ \alpha  \beta \nu} k^\alpha k^\beta X^\nu \, . 
\eea 
Thus, we have 
\bea
\label{nablaX}
k^\alpha \nabla_\alpha X^\mu \rvert_{\obs}  &=& \frac{X^\mu \rvert_{\emi} - X^\mu \rvert_{\obs}}{\Delta\lambda} -   R\UD{\mu}{ \alpha  \beta \nu} k^\alpha k^\beta  X^\nu   \rvert_{\obs} \,\Delta\lambda   
+ \mathcal{O}(\Delta\lambda^2) 
\eea
which we might contract project onto the screen space while using the boundary conditions \eqref{deviationini} to give 
\bea
\label{nablaX2}
p\UD{\sigma}{ \mu}\, k^\alpha \nabla_\alpha X^\mu \rvert_{\obs}  &=& p\UD{\sigma}{ \mu}  \rvert_{\obs}  \,\frac{E_\obs u^\mu \rvert_{\emi} }{ E_\emi \Delta\lambda} -   p\UD{\sigma}{\mu} \, R\UD{\mu}{ \alpha  \beta \nu}\, k^\alpha k^\beta  u^\nu   \rvert_{\obs} \,\Delta\lambda   + \mathcal{O}(\Delta\lambda^2) \, .  
\eea
Using \eqref{vector}, we write 
\bea
\label{uE} 
\hspace{-0.7cm} u^\mu /E \rvert_{\emi}  &=& u^\mu/E  \rvert_{\obs}  + \frac{1}{E}k^\alpha \nabla_\alpha u^\mu \rvert_{\obs}\, \Delta\lambda  +   u^\mu k^{\alpha} \nabla_\alpha (1/E)    \rvert_{\obs}\, \Delta\lambda   \nonumber \\ 
&&+ \left(  \frac{1}{2E} k^\alpha k^\beta ( \nabla_\alpha \! \nabla_\beta u^\mu  + R\UD{\mu}{ \alpha  \beta \nu} u^\nu  )     +   k^\alpha \nabla_\alpha (u^\mu) k^\beta \nabla_\beta (1/E)   + \frac{1}{2}  u^\mu k^\alpha k^\beta  \nabla_\alpha \! \nabla_\beta (1/E) \! \right) \! \Big\rvert_{\obs} \! \Delta\lambda^2    + \mathcal{O}(\Delta\lambda^3) , 
\eea  
which can be inserted into \eqref{nablaX2} to read 
\bea
\label{nablaX3}
\hspace{-0.6cm}  p\UD{\sigma}{ \mu} k^\alpha \nabla_\alpha X^\mu \rvert_{\obs}  &=& p\UD{\sigma}{\mu}  k^\alpha \nabla_\alpha u^\mu  \rvert_{\obs}  +  \left(\! E p\UD{\sigma}{ \mu}   k^\alpha \nabla_\alpha (u^\mu) k^\beta \nabla_\beta (1/E)    +  \frac{1}{2}  p\UD{\sigma}{\mu}  k^\alpha k^\beta ( \nabla_\alpha \! \nabla_\beta u^\mu  - R\UD{\mu}{ \alpha  \beta \nu} u^\nu  ) \! \right) \! \Big\rvert_{\obs} \! \Delta\lambda  \nonumber \\
&+& \mathcal{O}(\Delta\lambda^2) .  
\eea
We note that all terms but the first in \eqref{nablaX3} consistently vanish when $u^\mu$ is itself a deviation vector. 
We now use \eqref{nablaX3} and the identity 
 \bea
\label{Eevol}
k^\beta \nabla_\beta \frac{1}{E} \rvert_{\obs}  && =   \frac{1}{E^2} k^{\alpha}k^\beta \nabla_\alpha u_\beta  \rvert_{\obs} = e^{\alpha}e^\beta \nabla_\alpha u_\beta  \rvert_{\obs} - e^\beta a_\beta  \rvert_{\obs} 
\eea 
in \eqref{positiondrift} to obtain  
\bea
\label{positiondrift2}
 \hspace*{-0.8cm} &&\kappa^\sigma \rvert_{\obs}  =  e^\rho p\UD{ \sigma }{\mu }  \nabla_\rho u^\mu  \rvert_{\obs}  +  \Delta r \left[ \frac{1}{2} p\UD{ \sigma }{ \mu } u^\alpha u^\beta \nabla_\alpha \nabla_\beta u^\mu   + e^\alpha \left(- p\UD{ \sigma }{ \mu } a^\mu a_\alpha  -  p\UD{ \sigma }{ \mu } u^\beta \nabla_\alpha \nabla_\beta u^\mu + p\UD{ \sigma }{ \mu } R\UD{\mu}{   \beta \alpha \nu} u^\beta u^\nu   \right)  \right. \nonumber \\   
 \hspace*{-0.8cm} && \left. + e^\alpha e^\beta \left( p\UD{ \sigma }{\mu } a^\mu  \nabla_\alpha u_\beta +  a_\beta p\UD{ \sigma }{ \mu } \nabla_\alpha u^\mu + \frac{1}{2}  p\UD{ \sigma }{ \mu } \nabla_\alpha  \nabla_\beta u^\mu   - \frac{1}{2}  p\UD{ \sigma }{\mu }  R\UD{\mu}{  \beta \alpha \nu}  u^\nu  \right)    +  e^\alpha e^\beta e^\gamma  \left( - p\UD{ \sigma }{ \mu } \nabla_\gamma (u^\mu) \nabla_\alpha u_\beta  \right)   \right]_{\obs} \nonumber \\
 &&+  \mathcal{O}(\Delta r^2)  \, , 
\eea 
with $\Delta r \equiv  - E_\obs \Delta \lambda $.

\section{Series expansion of derivative of position drift}
Consider now the derivative of position drift: $k^\alpha \nabla_\alpha \kappa^\mu \rvert_{\obs}$. This quantity appears in higher order cosmography of the redshift drift signal, and is therefore convenient to express in terms of a multipole series expansion in $e^\mu$. 
We have 
 \bea
\label{kappaderiv}
 \hspace*{-0.3cm}  k^\alpha \nabla_\alpha \kappa^\mu \rvert_{\obs}   &&=  \frac{ \kappa^\mu \rvert_{\emi} - \kappa^\mu \rvert_{\obs}  }{\Delta \lambda}  + \mathcal{O}(\Delta\lambda) \nonumber \\ 
&& =    \frac{ -\frac{1}{E_\obs} \left(p\UD{ \mu }{ \nu }  k^\alpha \nabla_\alpha X^\nu  \rvert_{\emi} -p\UD{ \mu }{ \nu }  k^\alpha \nabla_\alpha X^\nu  \rvert_{\obs}  \right) }{\Delta \lambda} +  \frac{ p\UD{ \mu }{ \nu }  a^\nu \rvert_{\emi} - p\UD{ \mu }{ \nu }  a^\nu \rvert_{\obs}    }{\Delta \lambda} + \mathcal{O}(\Delta\lambda) \nonumber \\
&& =     \frac{ -\frac{1}{E_\obs} \left[p\UD{ \mu }{\nu } \rvert_{\obs}  (  k^\alpha \nabla_\alpha X^\nu  \rvert_{\emi} -   k^\alpha \nabla_\alpha X^\nu  \rvert_{\obs} ) +  k^\alpha \nabla_\alpha X^\nu  \rvert_{\obs} \Delta p\UD{ \mu }{ \nu }   \right]   }{\Delta \lambda}  +  \frac{ p\UD{ \mu }{ \nu } \rvert_{\obs}  (  a^\nu \rvert_{\emi} -  a^\nu \rvert_{\obs}) + a_\obs^\nu  \Delta p\UD{ \mu }{ \nu }     }{\Delta \lambda} + \mathcal{O}(\Delta\lambda) \nonumber \\ &&  
\eea 
with 
 \bea
\label{Pdelta}
\frac{\Delta p\UD{ \mu }{ \nu }}{\Delta \lambda}  \! \equiv \! \left[- k^\mu k_\nu  k^\alpha \nabla_\alpha \! \left( \! \frac{1}{E^2} \! \right)  + k^\mu k^\alpha \nabla_\alpha \! \left( \! \frac{u_\nu}{E} \! \right) + k_\nu k^\alpha \nabla_\alpha \! \left( \! \frac{u^\mu}{E} \! \right) \right]_\obs \, . 
\eea 
Using \eqref{Pdelta} and the geodesic deviation equation \eqref{Gdeviation}, we have 
 \bea
\label{kappaderiv}
 k^\alpha \nabla_\alpha \kappa^\mu \rvert_{\obs}    &=&    -\frac{1}{E_\obs} \left[p\UD{ \mu }{ \nu }    R\UD{\nu}{  \alpha  \beta \gamma} k^\alpha k^\beta X^\gamma  +   k^\mu p\UD{ \sigma }{ \nu }   k^\alpha \nabla_\alpha \! \left( \! \frac{u_\sigma}{E} \! \right)  k^\alpha \nabla_\alpha X^\nu     \right]_{\obs}    \nonumber \\ 
&&+   p\UD{ \mu }{ \nu }   k^\alpha \nabla_\alpha a^\nu \rvert_{\obs}  +  k^\mu   a^\nu    k^\alpha \nabla_\alpha \! \left( \! \frac{u_\nu}{E} \! \right)  \rvert_\obs  +  a^\nu k_\nu k^\alpha \nabla_\alpha \! \left( \! \frac{u^\mu}{E} \! \right)  \rvert_\obs  - k^\mu a^\nu k_\nu  k^\alpha \nabla_\alpha \! \left( \! \frac{1}{E^2} \! \right)  \rvert_\obs  + \mathcal{O}(\Delta\lambda)  \nonumber \\ 
&=&   -\frac{1}{E_\obs} \left[p\UD{ \mu }{\nu }    R\UD{\nu}{ \alpha  \beta \gamma} k^\alpha k^\beta u^\gamma  +  \frac{1}{E} k^\mu p\UD{ \sigma }{ \nu }   k^\alpha \nabla_\alpha (u_\sigma) k^\alpha \nabla_\alpha u^\nu     \right]_{\obs}    \nonumber \\ 
&&+   p\UD{ \mu }{ \nu }   k^\alpha \nabla_\alpha a^\nu \rvert_{\obs}  +  \frac{1}{E} k^\mu   a^\nu    k^\alpha \nabla_\alpha  u_\nu  \rvert_\obs   +  a^\nu k_\nu k^\alpha \nabla_\alpha \! \left( \! \frac{u^\mu}{E} \! \right)  \rvert_\obs   - k^\mu a^\nu k_\nu  k^\alpha \nabla_\alpha \! \left( \! \frac{1}{E^2} \! \right)  \rvert_\obs  + \mathcal{O}(\Delta\lambda) . 
\eea

\section{Proof that $\sigma_{\mu\nu}$, $\omega_{\mu\nu}$ and the observer's Newtonian acceleration can be obtained from the multipole decomposition of $\kappa^\mu$} \label{ap:vsh}

For the purpose of this Appendix, we will consider the celestial sphere as a unit sphere embedded in $\mathbf{R}^3$, with the components of vectors in $\mathbf{R}^3$ denoted by small Latin letters $i, j, k, \dots \in \{1, 2, 3\}$.
Recall that the three spatial components of the vector $e^i$, expressed in terms of the angular coordinates $\theta$ and $\varphi$, read
\bea
e^1 &=& \sin\theta\,\cos\varphi \nonumber \\
e^2 &=& \sin\theta\,\sin\varphi \nonumber \\
e^3 &=& \cos\theta. 
\eea
Functions $f(\theta, \varphi)$ defined on the celestial sphere $r=1$ can be identified with functions on the embedding space $\mathbf{R}^3 \setminus \left\{0\right\}$ if we simply interpret them in spherical coordinates  $(r, \theta, \varphi)$. In this case $f(\theta,\varphi)$ can be interpreted as a function constant on lines passing through the origin, given by $
\theta = \textrm{const}, \varphi = \textrm{const}$. 
We note that the 3-dimensional gradient $f^{,i}(\theta, \varphi)$ of such a function, evaluated at a point of the celestial sphere ($r=1$), is a vector tangent to this sphere and corresponds to the 2-dimensional gradient of that function on the celestial sphere.

One can verify the following formula for the derivatives of the 3 functions $e^j$:
\bea
\partial_j e^i = \delta\UD{i}{j} - e^i\,e_j \equiv p\UD{i}{j}.\label{eq:gradiend of e}
\eea

Any vector field tangent to a 2-sphere can be uniquely expanded in terms of the poloidal and toroidal spherical harmonics. The former corresponds to the gradient of a function on the sphere, the latter to the rotation of the gradient of a function, i.e.
\bea
\kappa^i = F^{,i} + \epsilon\UD{ij}{k}\,G_{,j}\,e^k,
\eea
with $F$ and $G$ denoting the functions, $\epsilon_{ijk}$ being the antisymmetric tensor and $\epsilon_{ijk}\,e^k$ representing the area 2-form on a sphere.  
If we have the multipole decompositions of $F$ and it $G$, we can explore this identity to get the multipole decompostion of $\kappa^i$. 

We will now show that the acceleration term $\dot v_\mu$ corresponds to the dipole of $F$, $\sigma_{\mu\nu}$ to the quadrupole of $F$ and $\omega_{\mu\nu}$ to the dipole of $G$. Assume that we have
\bea
F &=& F_i\,e^i + \frac{1}{2}F_{\langle ij \rangle}\,e^i\,e^j \nonumber\\
G &=& G_i\,e^i. 
\eea
We calculate their gradients with the help of \eqref{eq:gradiend of e}:
\bea
F^{k} &=& F_i\,p^{ik} + F_{\langle ij\rangle}\,e^j\,p^{ik} \nonumber\\
G_{,k} &=& G_{i}\,p\UD{i}{k} 
\eea
and it follows that
\bea
\kappa^k = F_i\,p^{ik} + F_{\langle ij\rangle}\,e^j\,p^{ik} + \epsilon\UD{kl}{m}\,G_b\,p\UD{b}{l}\,e^m. \label{eq:kappa VSH expansion 1}
\eea
Note that $\epsilon\UD{kl}{m}\,p\UD{b}{l}\,e^m = \epsilon\UD{kb}{m}\,e^m = p\UD{k}{c}\,\epsilon\UD{cb}{m}\,e^m$ because
$\epsilon\UD{kb}{m}\,e^m$ is orthogonal to $e^i$ in both $k$ and $b$ indices. We can therefore rewrite \eqref{eq:kappa VSH expansion 1} as
\bea
\kappa^k = F_i\,p^{ik} + F_{\langle ij\rangle}\,e^j\,p^{ik} + \,p\UD{k}{c}\,\epsilon\UD{cl}{m}\,G_l\,e^m.  
\eea
Comparing this expression with \eqref{pdseries}-\eqref{pdcoefficients} we see that we can identify $F_{\langle ij \rangle} = \sigma_{ij}$ and $\epsilon\UD{cl}{m}\,G_l= \omega\UD{c}{m}$. Note that the second relation is perfectly invertible, i.e. $G_l = -\frac{1}{2}\epsilon_{ijl}\,\omega^{ij}$.
Since all components of $F_{\langle ij \rangle}$ and $G_i$ can be recovered from the vector field $\kappa^i(e^j)$, it follows that all components both $\sigma_{ij}$ and $\omega_{ij}$ can be recovered from the all-sky measurements of $\kappa^i$.

Moreover, the local Newtonian 3-acceleration $\dot v_j$, not taken into account in this paper, enters the position drift $\kappa^k$ via the aberration drift term
$\dot v_j\,p^{kj}$ \cite{Marcori:2018cwn}.  This allows us to identify the acceleration with the poloidal dipole: $F_i = \dot v_i$. Again, it follows that all of its components can be recovered from $\kappa^i$ independently of the shear and vorticity.

\twocolumngrid 

\bibliographystyle{mnras_unsrt}
\bibliography{refs}

\begin{thebibliography}{}
\makeatletter
\relax
\def\mn@urlcharsother{\let\do\@makeother \do\$\do\&\do\#\do\^\do\_\do\%\do\~}
\def\mn@doi{\begingroup\mn@urlcharsother \@ifnextchar [ {\mn@doi@}
  {\mn@doi@[]}}
\def\mn@doi@[#1]#2{\def\@tempa{#1}\ifx\@tempa\@empty \href
  {http://dx.doi.org/#2} {doi:#2}\else \href {http://dx.doi.org/#2} {#1}\fi
  \endgroup}
\def\mn@eprint#1#2{\mn@eprint@#1:#2::\@nil}
\def\mn@eprint@arXiv#1{\href {http://arxiv.org/abs/#1} {{\tt arXiv:#1}}}
\def\mn@eprint@dblp#1{\href {http://dblp.uni-trier.de/rec/bibtex/#1.xml}
  {dblp:#1}}
\def\mn@eprint@#1:#2:#3:#4\@nil{\def\@tempa {#1}\def\@tempb {#2}\def\@tempc
  {#3}\ifx \@tempc \@empty \let \@tempc \@tempb \let \@tempb \@tempa \fi \ifx
  \@tempb \@empty \def\@tempb {arXiv}\fi \@ifundefined
  {mn@eprint@\@tempb}{\@tempb:\@tempc}{\expandafter \expandafter \csname
  mn@eprint@\@tempb\endcsname \expandafter{\@tempc}}}

\bibitem[\protect\citeauthoryear{{Sandage}}{{Sandage}}{1962}]{1962ApJ...136..319S}
{Sandage} A.,  1962, \mn@doi [Astrophys. J.] {10.1086/147385}, \href
  {https://ui.adsabs.harvard.edu/abs/1962ApJ...136..319S} {136, 319}

\bibitem[\protect\citeauthoryear{{McVittie}}{{McVittie}}{1962}]{1962ApJ...136..334M}
{McVittie} G.~C.,  1962, Astrophys. J., \href
  {https://ui.adsabs.harvard.edu/abs/1962ApJ...136..334M} {136, 334}

\bibitem[\protect\citeauthoryear{{Loeb}}{{Loeb}}{1998}]{Loeb:1998bu}
{Loeb} A.,  1998, {Astrophys. J. Lett.}, 499, L111

\bibitem[\protect\citeauthoryear{Martins}{Martins}{2019}]{Martins:2019gxw}
Martins C. J. A.~P.,  2019, in {IAU Symposium 347}: {Early Science with ELTs
  (EASE)}.  (\mn@eprint {arXiv} {1902.01783})

\bibitem[\protect\citeauthoryear{Maartens, Abdalla, Jarvis  \& Santos}{Maartens
  et~al.}{2015}]{Maartens:2015mra}
Maartens R.,  Abdalla F.~B.,  Jarvis M.,   Santos M.~G.,  2015, \mn@doi [PoS]
  {10.22323/1.215.0016}, AASKA14, 016

\bibitem[\protect\citeauthoryear{Quercellini, Quartin  \& Amendola}{Quercellini
  et~al.}{2009}]{Quercellini:2008ty}
Quercellini C.,  Quartin M.,   Amendola L.,  2009, \mn@doi [Phys. Rev. Lett.]
  {10.1103/PhysRevLett.102.151302}, 102, 151302

\bibitem[\protect\citeauthoryear{Krasinski \& Bolejko}{Krasinski \&
  Bolejko}{2011}]{Krasinski:2010rc}
Krasinski A.,  Bolejko K.,  2011, \mn@doi [Phys. Rev. D]
  {10.1103/PhysRevD.83.083503}, 83, 083503

\bibitem[\protect\citeauthoryear{Quercellini, Amendola, Balbi, Cabella  \&
  Quartin}{Quercellini et~al.}{2012}]{Quercellini:2010zr}
Quercellini C.,  Amendola L.,  Balbi A.,  Cabella P.,   Quartin M.,  2012,
  \mn@doi [Phys. Rept.] {10.1016/j.physrep.2012.09.002}, 521, 95

\bibitem[\protect\citeauthoryear{{Gaia Collaboration} et~al.,}{{Gaia
  Collaboration} et~al.}{2018}]{2018A&A...616A..14G}
{Gaia Collaboration} et~al., 2018, \mn@doi [\aap]
  {10.1051/0004-6361/201832916}, \href
  {https://ui.adsabs.harvard.edu/abs/2018A&A...616A..14G} {616, A14}

\bibitem[\protect\citeauthoryear{{Gaia Collaboration} et~al.,}{{Gaia
  Collaboration} et~al.}{2021a}]{2021A&A...649A...9G}
{Gaia Collaboration} et~al., 2021a, \mn@doi [\aap]
  {10.1051/0004-6361/202039734}, \href
  {https://ui.adsabs.harvard.edu/abs/2021A&A...649A...9G} {649, A9}

\bibitem[\protect\citeauthoryear{Korzy\'nski \& Kopi\'nski}{Korzy\'nski \&
  Kopi\'nski}{2018}]{Korzynski:2017nas}
Korzy\'nski M.,  Kopi\'nski J.,  2018, \mn@doi [JCAP]
  {10.1088/1475-7516/2018/03/012}, 03, 012

\bibitem[\protect\citeauthoryear{Liske et~al.}{Liske
  et~al.}{2008}]{Liske:2008ph}
Liske J.,  et~al., 2008, \mn@doi [Mon. Not. Roy. Astron. Soc.]
  {10.1111/j.1365-2966.2008.13090.x}, 386, 1192

\bibitem[\protect\citeauthoryear{R\"as\"anen}{R\"as\"anen}{2014}]{Rasanen:2013swa}
R\"as\"anen S.,  2014, \mn@doi [JCAP] {10.1088/1475-7516/2014/03/035}, 03, 035

\bibitem[\protect\citeauthoryear{Marcori, Pitrou, Uzan  \& Pereira}{Marcori
  et~al.}{2018}]{Marcori:2018cwn}
Marcori O.~H.,  Pitrou C.,  Uzan J.-P.,   Pereira T.~S.,  2018, \mn@doi [Phys.
  Rev. D] {10.1103/PhysRevD.98.023517}, 98, 023517

\bibitem[\protect\citeauthoryear{Bessa, Durrer  \& Stock}{Bessa
  et~al.}{2023}]{Bessa:2023qrr}
Bessa P.,  Durrer R.,   Stock D.,  2023, \mn@doi [JCAP]
  {10.1088/1475-7516/2023/11/093}, 11, 093

\bibitem[\protect\citeauthoryear{Fleury, Pitrou  \& Uzan}{Fleury
  et~al.}{2015}]{Fleury:2014rea}
Fleury P.,  Pitrou C.,   Uzan J.-P.,  2015, \mn@doi [Phys. Rev. D]
  {10.1103/PhysRevD.91.043511}, 91, 043511

\bibitem[\protect\citeauthoryear{Koksbang}{Koksbang}{2020}]{Koksbang:2020zej}
Koksbang S.~M.,  2020, \mn@doi [\MNRAS] {10.1093/mnrasl/slaa146}, 498, L135

\bibitem[\protect\citeauthoryear{{Quartin} \& {Amendola}}{{Quartin} \&
  {Amendola}}{2010}]{2010PhRvD..81d3522Q}
{Quartin} M.,  {Amendola} L.,  2010, \mn@doi [\prd]
  {10.1103/PhysRevD.81.043522}, \href
  {https://ui.adsabs.harvard.edu/abs/2010PhRvD..81d3522Q} {81, 043522}

\bibitem[\protect\citeauthoryear{Koksbang}{Koksbang}{2023a}]{Koksbang:2023wez}
Koksbang S.~M.,  2023a, \mn@doi [Phys. Rev. Lett.]
  {10.1103/PhysRevLett.130.201003}, 130, 201003

\bibitem[\protect\citeauthoryear{Koksbang}{Koksbang}{2023b}]{Koksbang:2023tun}
Koksbang S.~M.,  2023b, \mn@doi [Phys. Rev. D] {10.1103/PhysRevD.107.063544},
  107, 063544

\bibitem[\protect\citeauthoryear{Koksbang, Heinesen  \& Macpherson}{Koksbang
  et~al.}{2024}]{Koksbang:2024xfr}
Koksbang S.~M.,  Heinesen A.,   Macpherson H.~J.,  2024, {Redshift drift in a
  universe with structure III: Numerical relativity} (\mn@eprint {arXiv}
  {2404.06242})

\bibitem[\protect\citeauthoryear{Hellaby \& Walters}{Hellaby \&
  Walters}{2018}]{Hellaby:2017soj}
Hellaby C.,  Walters A.,  2018, \mn@doi [JCAP] {10.1088/1475-7516/2018/02/015},
  02, 015

\bibitem[\protect\citeauthoryear{Korzy\'nski \& Villa}{Korzy\'nski \&
  Villa}{2020}]{Korzynski:2019oal}
Korzy\'nski M.,  Villa E.,  2020, \mn@doi [Phys. Rev. D]
  {10.1103/PhysRevD.101.063506}, 101, 063506

\bibitem[\protect\citeauthoryear{Heinesen}{Heinesen}{2021a}]{Heinesen:2020pms}
Heinesen A.,  2021a, \mn@doi [Phys. Rev. D] {10.1103/PhysRevD.103.023537}, 103,
  023537

\bibitem[\protect\citeauthoryear{Heinesen}{Heinesen}{2021b}]{Heinesen:2021nrc}
Heinesen A.,  2021b, \mn@doi [Phys. Rev. D] {10.1103/PhysRevD.103.L081302},
  103, L081302

\bibitem[\protect\citeauthoryear{Grasso \& Villa}{Grasso \&
  Villa}{2022}]{Grasso:2021iwq}
Grasso M.,  Villa E.,  2022, \mn@doi [Class. Quant. Grav.]
  {10.1088/1361-6382/ac35aa}, 39, 015011

\bibitem[\protect\citeauthoryear{Grasso, Villa, Korzy\ifmmode~\acute{n}\else
  \'{n}\fi{}ski  \& Matarrese}{Grasso et~al.}{2021}]{PhysRevD.104.043508}
Grasso M.,  Villa E.,  Korzy\ifmmode~\acute{n}\else \'{n}\fi{}ski M.,
  Matarrese S.,  2021, \mn@doi [Phys. Rev. D] {10.1103/PhysRevD.104.043508},
  104, 043508

\bibitem[\protect\citeauthoryear{Heinesen}{Heinesen}{2021c}]{Heinesen:2021qnl}
Heinesen A.,  2021c, \mn@doi [Phys. Rev. D] {10.1103/PhysRevD.104.123527}, 104,
  123527

\bibitem[\protect\citeauthoryear{Umeh}{Umeh}{2013}]{Umeh:2013UCT}
Umeh O.,  2013, PhD thesis, University of Cape Town, Faculty of Science,
  Department of Mathematics and Applied Mathematics,
  https://open.uct.ac.za/handle/11427/4938

\bibitem[\protect\citeauthoryear{Clarkson \& Umeh}{Clarkson \&
  Umeh}{2011}]{Clarkson:2011uk}
Clarkson C.,  Umeh O.,  2011, \mn@doi [Class. Quant. Grav.]
  {10.1088/0264-9381/28/16/164010}, 28, 164010

\bibitem[\protect\citeauthoryear{Heinesen}{Heinesen}{2021d}]{Heinesen:2020bej}
Heinesen A.,  2021d, \mn@doi [JCAP] {10.1088/1475-7516/2021/05/008}, 05, 008

\bibitem[\protect\citeauthoryear{Maartens, Santiago, Clarkson, Kalbouneh  \&
  Marinoni}{Maartens et~al.}{2023}]{Maartens:2023tib}
Maartens R.,  Santiago J.,  Clarkson C.,  Kalbouneh B.,   Marinoni C.,  2023.
  (\mn@eprint {arXiv} {2312.09875})

\bibitem[\protect\citeauthoryear{Spencer}{Spencer}{1970}]{Spencer:1970}
Spencer A.,  1970, \mn@doi [International Journal of Engineering Science]
  {https://doi.org/10.1016/0020-7225(70)90024-8}, 8, 475

\bibitem[\protect\citeauthoryear{Grasso, Korzy\'nski  \& Serbenta}{Grasso
  et~al.}{2019}]{Grasso:2018mei}
Grasso M.,  Korzy\'nski M.,   Serbenta J.,  2019, \mn@doi [Phys. Rev. D]
  {10.1103/PhysRevD.99.064038}, 99, 064038

\bibitem[\protect\citeauthoryear{Inoue, Komatsu, Aoki, Chiba, Misawa  \&
  Usuda}{Inoue et~al.}{2019}]{inoue2019}
Inoue T.,  Komatsu E.,  Aoki W.,  Chiba T.,  Misawa T.,   Usuda T.,  2019,
  \mn@doi [Publications of the Astronomical Society of Japan]
  {10.1093/pasj/psz131}, 72, L1

\bibitem[\protect\citeauthoryear{Lobo, Mimoso  \& Visser}{Lobo
  et~al.}{2020}]{Lobo:2020hcz}
Lobo F. S.~N.,  Mimoso J.~P.,   Visser M.,  2020, \mn@doi [JCAP]
  {10.1088/1475-7516/2020/04/043}, 04, 043

\bibitem[\protect\citeauthoryear{{Mignard} \& {Klioner}}{{Mignard} \&
  {Klioner}}{2012}]{2012A&A...547A..59M}
{Mignard} F.,  {Klioner} S.,  2012, \mn@doi [\aap]
  {10.1051/0004-6361/201219927}, \href
  {https://ui.adsabs.harvard.edu/abs/2012A&A...547A..59M} {547, A59}

\bibitem[\protect\citeauthoryear{Barrera, Estevez  \& Giraldo}{Barrera
  et~al.}{1985}]{RGBarrera_1985}
Barrera R.~G.,  Estevez G.~A.,   Giraldo J.,  1985, \mn@doi [European Journal
  of Physics] {10.1088/0143-0807/6/4/014}, 6, 287

\bibitem[\protect\citeauthoryear{{Xu, M. H.}, {Wang, G. L.}  \& {Zhao,
  M.}}{{Xu, M. H.} et~al.}{2012}]{Xu2012}
{Xu, M. H.} {Wang, G. L.}  {Zhao, M.} 2012, \mn@doi [A&A]
  {10.1051/0004-6361/201219593}, 544, A135

\bibitem[\protect\citeauthoryear{{Titov, O.}, {Lambert, S. B.}  \& {Gontier,
  A.-M.}}{{Titov, O.} et~al.}{2011}]{titov2011}
{Titov, O.} {Lambert, S. B.}  {Gontier, A.-M.} 2011, \mn@doi [A&A]
  {10.1051/0004-6361/201015718}, 529, A91

\bibitem[\protect\citeauthoryear{{Gaia Collaboration} et~al.,}{{Gaia
  Collaboration} et~al.}{2021b}]{gaia2021}
{Gaia Collaboration} et~al., 2021b, \mn@doi [A&A]
  {10.1051/0004-6361/202039734}, 649, A9

\bibitem[\protect\citeauthoryear{Macpherson \& Heinesen}{Macpherson \&
  Heinesen}{2021}]{Macpherson:2021gbh}
Macpherson H.~J.,  Heinesen A.,  2021, \mn@doi [Phys. Rev. D]
  {10.1103/PhysRevD.104.023525}, 104, 023525

\bibitem[\protect\citeauthoryear{Heinesen \& Macpherson}{Heinesen \&
  Macpherson}{2022}]{Heinesen:2021azp}
Heinesen A.,  Macpherson H.~J.,  2022, \mn@doi [JCAP]
  {10.1088/1475-7516/2022/03/057}, 03, 057

\bibitem[\protect\citeauthoryear{Dhawan, Borderies, Macpherson  \&
  Heinesen}{Dhawan et~al.}{2023}]{Dhawan:2022lze}
Dhawan S.,  Borderies A.,  Macpherson H.~J.,   Heinesen A.,  2023, \mn@doi
  [Mon. Not. Roy. Astron. Soc.] {10.1093/mnras/stac3812}, 519, 4841

\bibitem[\protect\citeauthoryear{Macpherson}{Macpherson}{2023}]{Macpherson:2022eve}
Macpherson H.~J.,  2023, \mn@doi [JCAP] {10.1088/1475-7516/2023/03/019}, 03,
  019

\bibitem[\protect\citeauthoryear{Koksbang \& Heinesen}{Koksbang \&
  Heinesen}{2022}]{Koksbang:2022upf}
Koksbang S.~M.,  Heinesen A.,  2022, \mn@doi [Phys. Rev. D]
  {10.1103/PhysRevD.106.043501}, 106, 043501

\bibitem[\protect\citeauthoryear{Cowell, Dhawan  \& Macpherson}{Cowell
  et~al.}{2022}]{Cowell:2022ehf}
Cowell J.~A.,  Dhawan S.,   Macpherson H.~J.,  2022.  (\mn@eprint {arXiv}
  {2212.13569})

\bibitem[\protect\citeauthoryear{Martín-García et~al.,}{Martín-García
  et~al.}{2024}]{xAct}
Martín-García J.~M.,  et~al., GPL 2002-2024, xAct: Efficient tensor computer
  algebra for the Wolfram Language, \url {http://www.xact.es}

\bibitem[\protect\citeauthoryear{Martín-García}{Martín-García}{2024}]{xTensor}
Martín-García J.~M.,  GPL 2002-2024, xTensor: Fast abstract tensor computer
  algebra, \url {http://www.xact.es/xTensor}

\bibitem[\protect\citeauthoryear{Martín-García}{Martín-García}{2008}]{MARTINGARCIA2008597}
Martín-García J.~M.,  2008, \mn@doi [Computer Physics Communications]
  {https://doi.org/10.1016/j.cpc.2008.05.009}, 179, 597

\bibitem[\protect\citeauthoryear{Adamek, Clarkson, Durrer, Heinesen, Kunz  \&
  Macpherson}{Adamek et~al.}{2024}]{adamek2024cosmography}
Adamek J.,  Clarkson C.,  Durrer R.,  Heinesen A.,  Kunz M.,   Macpherson
  H.~J.,  2024, Towards Cosmography of the Local Universe (\mn@eprint {arXiv}
  {2402.12165})

\makeatother
\end{thebibliography}

\end{document}